\documentclass[fleqn,usenatbib]{mnras}

\usepackage{newtxtext,newtxmath}
\usepackage[T1]{fontenc}

\DeclareRobustCommand{\VAN}[3]{#2}
\let\VANthebibliography\thebibliography
\def\thebibliography{\DeclareRobustCommand{\VAN}[3]{##3}\VANthebibliography}

\usepackage{graphicx}	% Including figure files
\usepackage{amsmath}	% Advanced maths commands

\title[Q2237+0305 ``planets'']{Free-floating ``planets'' in the macrolensed quasar Q2237+0305}

\author[A.V. Tuntsov et al]{
Artem V. Tuntsov,$^{1}$\thanks{E-mail: artem.tuntsov@manlyastrophysics.org}
Geraint F. Lewis,$^{2}$
Mark A. Walker$^{1}$
\\
$^{1}$Manly Astrophysics, 15/41-42 East Esplanade, Manly, NSW 2095, Australia\\
$^{2}$Sydney Institute for Astronomy, School of Physics A28, The University of Sydney, NSW 2006, Australia
}
\date{Accepted 2024 January 08. Received 2024 January 08; in original form 2023 April 27}

\pubyear{2023}

\begin{document}
\label{firstpage}
\pagerange{\pageref{firstpage}--\pageref{lastpage}}
\maketitle

\newcommand{\vdag}{(v)^\dagger}
\newcommand\aastex{AAS\TeX}
\newcommand\latex{La\TeX}
\newcommand{\source}{{Q2237+0305}}
\newcommand{\htwo}{H$_2$}

%% Tells LaTeX to search for image files in the 
%% current directory as well as in the figures/ folder.
\graphicspath{{./}{figures/}}

\begin{abstract}

It has been claimed that the variability of field quasars resembles gravitational lensing by a large cosmological population of free-floating planets with mass $\sim 10\, {\rm M}_\oplus$. But Galactic photometric monitoring experiments, on the other hand, exclude a large population of such planetary-mass gravitational lenses. These apparently contradictory pieces of evidence can be reconciled if the objects under consideration have a mean column-density that lies between the critical column-densities for gravitational lensing in these two contexts. Dark matter in that form is known to be weakly collisional, so that a core develops in galaxy halo density profiles, and a preferred model has already been established. Here we consider what such a model implies for \source, which is the best-studied example of a quasar that is strongly lensed by an intervening galaxy. We construct microlensing magnification maps appropriate to the four macro-images of the quasar --- all of which are seen through the bulge of the galaxy. Each of these maps exhibits a caustic network arising from the stars, plus many small, isolated caustics arising from the free-floating ``planets'' in the lens galaxy. The ``planets'' have little influence on the magnification histograms but a large effect on the statistics of the magnification gradients. We compare our predictions to the published OGLE photometry of \source\ and find that these data are consistent with the presence of the hypothetical ``planets''. However, the evidence is relatively weak because the OGLE dataset is not well suited to testing our predictions and requires low-pass filtering for this application. New data from a large, space-based telescope are desirable to address this issue.
\end{abstract}

\newcommand{\md}{\mathrm{d}}
\newcommand{\vvec}{{\bf v}}
\newcommand{\uvec}{{\bf u}}
\newcommand{\rate}{{\cal R}}
\newcommand{\drm}{{\rm d}}
\newcommand\be{\begin{equation}}
\newcommand\ee{\end{equation}}


\section{Introduction}

Dark matter, first inferred by Fritz \citet[][republication in English translation of the 1933 German original]{2009GReGr..41..207Z}, has been the focus of a great deal of attention over the last few decades \citep{1987ARA&A..25..425T,2017NatAs...1E..59D}. It appears to be the main form of matter in the universe, but its fundamental nature is still unclear. Models of galaxy formation have demonstrated considerable success when the model universe starts out with dark matter predominantly in the form of non-relativistic elementary particles having no electromagnetic interaction \citep[e.g.][]{1985ApJ...292..371D,1999coph.book.....P}. However, there has been no firm detection of a suitable elementary particle in the laboratory \citep[e.g.][]{2010ARA&A..48..495F,2018ARNPS..68..429B}. Although other forms of dark matter might lack such an attractive theoretical framework one can nevertheless look for observational manifestations of any kind; and searches for the gravitational lensing signature of macroscopic lumps of dark matter have been informative, as follows.

\cite{1986ApJ...304....1P} focused the attention of the astronomical community on the use of gravitational lensing to detect lumps of dark matter in the Galactic halo. The expected optical depth for this process is very small, so a large number of background stars must be monitored in order to make progress, but with a clear target to aim at several groups took up the challenge \citep[][]{1993Natur.365..621A,1993Natur.365..623A,1993AcA....43..289U}. Because of the short time-scales that go hand-in-hand with a small Einstein ring radius, a given optical depth leads to many more expected events if it resides in low mass lumps. But very few short duration events were detected and a quasi-spherical Galactic dark halo consisting predominantly of planetary-mass compact objects is now ruled out \citep{1998ApJ...499L...9A,2007A&A...469..387T,2011MNRAS.413..493W,2014ApJ...786..158G, 2022A&A...664A.106B}. More recently some short duration microlensing events have been detected in photometric monitoring of Galactic bulge stars, and interpreted as free-floating planets \citep{2011Natur.473..349S}, but that population has been shown to be much smaller than first thought \citep{2017Natur.548..183M, 2023AJ....166..108S}, and is dynamically insignificant. Constraints are also available for compact lenses in the planetary and sub-planetary mass range from high-cadence Subaru monitoring of M31 \citep{2023arXiv230813593D}.

The non-detection of planetary-mass lenses in the Galactic halo means that planetary-mass black holes are now an unattractive dark matter candidate. However, before the Galactic photometric monitoring experiments had delivered their results it had already been proposed by \cite{1993Natur.366..242H,1996MNRAS.278..787H} that a large, cosmological population of planetary-mass black holes could be responsible for the optical variability that is observed in field quasars. And, separately, \cite{1996ApJ...464..125S} had argued for a large population of what he called ``rogue planets'', on the basis of a possible microlensing signal in the macrolensed quasar Q0957+561. These claims were startling when they were first made, but neither one has been categorically refuted. There is, however, considerable ambiguity because of uncertainty about the size of quasar emission regions. If the source radius is as small as $\sim3\times10^{14}\,{\rm cm}$ then the low amplitude of any microlensing fluctuations in Q0957+561 excludes Schild's rogue planets making up all of the dark matter in the lens galaxy, whereas they are allowed if the source is as large as $\sim3\times10^{15}\,{\rm cm}$ \citep{2000A&A...362L..37W}. Additional ambiguity arises from the unknown transverse velocities of the source and microlenses. For example: assuming stellar mass microlenses and a source radius as large as $\sim4\times10^{16}\,{\rm cm}$, \cite{2003ApJ...584..657S} attributed the rapid, low-amplitude microlensing fluctuations of the macrolensed quasar HE~1104$-$1805 to relativistic transverse motion of the source.

The scenario advocated by \cite{1993Natur.366..242H} was analysed by \cite{1993A&A...279....1S}, who used quasar photometry from \cite{1993MNRAS.260..202H} to limit the possible cosmological abundance of microlenses, as a function of lens mass --- the point being that large flux excursions are observed to be very rare in practice, whereas massive lenses should sometimes introduce high magnifications. However, the \cite{1993A&A...279....1S} study did also show that a large fraction of the cosmological critical density could be present in lenses of mass $\la 10^{-4}\,{\rm M_\odot}$ without generating any conflict with the observed quasar lightcurves.
\cite{2003A&A...399...23Z} and \cite{2003A&A...408...17Z} independently reassessed the lensing interpretation of field quasar variability, finding it to be viable\footnote{Although, they argued, not the whole story: low redshift AGN vary intrinsically, so presumably quasars do also at some level.} when taken in isolation. \cite{2003ApJ...589..844W} investigated the effect of  nanolenses on gamma-ray bursts but found the available data could not significantly constrain their abundance.

As described above, the Galactic photometric monitoring experiments exclude the possibility that the Galactic dark matter resides in either planets or planetary-mass black holes; but objects that have a peak column-density $\Sigma_0 \la 10^5\;{\rm g\, cm^{-2}}$ do not automatically violate the Galactic constraints because they're not strong gravitational lenses in that context. Such objects do have a photometric signature when they occult Galactic stars, both from gas lensing \citep{1998ApJ...509L..41D,2001ApJ...547..207R} and dust extinction \citep{2002MNRAS.332L..29K,2003ApJ...589..281D}. However, the  form of the resulting light-curves is at present unknown and consequently a population of this type is difficult to constrain: we do not know what signature to look for. Furthermore there is a fundamental difficulty in using {\it any\/} type of optical occultation signal to constrain a population of dusty objects in the Galaxy, because dust extinction creates an anticorrelation between foreground clouds and detectable background stars.\footnote{Thus motivating work at longer wavelengths where the extinction is negligible (Suvorov et al., in prep).}

With those points in mind \cite{2022MNRAS.513.2491T} revisited the possibility of a large cosmological population of planetary-mass lenses, giving attention specifically to objects that are weak gravitational lenses when placed in the Galaxy, but strong at cosmological distances. Although these characteristics are not known to correspond to any form of matter that has been directly observed, so it is debatable whether such objects exist at all in our universe, it is worth mentioning that there are at least some theoretical counterparts in the structural models of \htwo\ snow clouds presented by \cite{2019ApJ...881...69W}.

For the larger column-densities in the range studied by \cite{2022MNRAS.513.2491T}, the magnification statistics for a given lens mass were found to be largely independent of $\Sigma_0$. That result is easily understood: once the objects become very strong gravitational lenses the dominant images form far from the centre (``outside the cloud''), where the densities of gas and dust are too small to have a significant effect on the measured flux. In that regime the constraints reported by \cite{2022MNRAS.513.2491T} parallel those previously reported by \cite{1993A&A...279....1S}, as indeed they should because the new simulations followed essentially the same approach as the original. In that respect the main difference between the two works is the $3\times$ larger source size adopted by \cite{2022MNRAS.513.2491T}. However, \cite{2022MNRAS.513.2491T} also undertook simulations for lens masses lower than \cite{1993A&A...279....1S} considered, and by doing so they were able to demonstrate that the mass range $10^{-4.5\pm0.5}\;{\rm M_\odot}$ is preferred, in the sense that the corresponding magnification statistics can quite closely resemble those of the \cite{1993MNRAS.260..202H} data.

Unfortunately, even a model that nicely reproduces the \cite{1993MNRAS.260..202H} data cannot be considered strong evidence regarding the nature of the dark matter, because we cannot be sure that field quasars really are varying primarily as a result of gravitational lensing rather than varying intrinsically. To address that issue we need instead to examine the variability of quasars that are macro-lensed --- i.e. strongly lensed by a foreground galaxy. For such systems, any variations that are intrinsic to the source show up in all of the macro-images, whereas the lumpiness in the mass distribution within the lens galaxy leads to different microlensing variations for each of those images  \citep{1981ApJ...243..140G}. This paper therefore addresses the question of what behaviour is predicted for a macrolensed quasar when the dark matter in the lensing galaxy is made up predominantly of the type of objects preferred by \cite{2022MNRAS.513.2491T}, and how those predictions compare to observations. For simplicity, in this paper we illustrate the expected nanolensing signals using a model in which \emph{all} of the ``missing mass'' inferred from dynamics is in the form of dense gas clouds, with no significant contribution from non-baryonic matter. As noted earlier, this sort of model is not preferred by cosmologists; nevertheless it forms a clear limiting case from which useful insights can be obtained.

Many examples of macrolensed quasars are now known \citep[e.g.][]{2010ARA&A..48...87T}. In this paper we focus on just one system, \source, in which the lens galaxy has an exceptionally low redshift \citep[$z=0.0394$,][]{1985AJ.....90..691H}, making it well suited to our application (for reasons that are discussed later, in \S6.2). Optical monitoring of \source\ has been undertaken by the OGLE team for more than a decade, and the resulting, high quality data have been placed in the public domain \citep{2000ApJ...529...88W,2006AcA....56..293U}; we make use of those data in this paper. Many analyses of \source\ lightcurves have been undertaken previously \citep[e.g.][]{2004ApJ...605...58K,2010ApJ...712..658P}, including work that gave attention to planetary-mass objects \citep[][who ruled out Jupiter-mass compact objects as a significant contribution to the mass of the bulge of the lens galaxy]{2000MNRAS.315...51W}.

There are several motivations to revisit the constraints on planetary-mass lumps in \source: first, we have a preferred mass range to focus on, so the model predictions are correspondingly specific; secondly, in the time since \cite{2000MNRAS.315...51W} was published the total duration of the OGLE lightcurves for \source\ has increased five-fold; thirdly, the properties of the stellar bulge of the lens galaxy are now accurately determined \citep{2010ApJ...719.1481V}; and, finally, the planetary-mass objects under consideration here are much less compact than was assumed in previous analyses, with important implications for the lensing model. The point there is that dark halos made up of low column-density objects are weakly collisional and develop a core in their density profile \citep{1999MNRAS.308..551W}. Moreover a preferred column-density for the individual dark matter lumps has already been determined from consideration of their collisional properties \citep{1999MNRAS.308..551W}, and that in turn fixes a specific value for the central, nanolensing contribution to the optical depth of the lens galaxy --- as described in \S\ref{subsection:coreddarkhalo}.

The finite column-density of the individual dark matter lumps also moderates the gravitational deflection angle, for light that passes through the lump itself; however, that is expected to have little influence on the magnification maps of \source, for two reasons. First is that the preferred column of the individual lumps, $\langle\Sigma\rangle \simeq 140\;{\rm g\, cm^{-2}}$ (see \S2.1), is more than $60\times$ larger than the critical surface-density for strong gravitational lensing in \source. In other words the individual lumps are very much smaller than their own Einstein rings. And secondly, as discussed in \S2.1, the central nanolensing optical depth of the lens galaxy is expected to be of order 0.1. Thus the finite column-density of the individual lumps makes no difference to 99.9\% of the rays (99.9\% of the area in the lens plane), and a point-mass approximation for the individual deflection angles is appropriate almost everywhere. Furthermore, the small fraction of rays for which the approximation is poor are themselves mostly not important, because they are typically strongly demagnified (``over-focused''). We therefore employ the point-mass deflection angle approximation in our ray-shooting calculations in \S3.

Because we are using the point-mass approximation for the deflection angle introduced by a planetary-mass lump, the results of the lensing calculations presented here are identical to what would be obtained if the same total mass were placed in planets rather than dense gas clouds. Consequently, for the remainder of this paper we will usually refer to the planetary-mass lumps as ``planets'', with the inverted commas serving to remind readers that they are not actual planets.

This paper is structured as follows. In \S\ref{section:macromodel} we present our macro model of the \source\ lens, describing both the stellar component and the cored dark matter distribution. Then in \S\ref{section:simulations} we present our micro/nanolensing simulations. We undertake two distinct sets of calculations: one that includes both stars and the planetary-mass lumps of dark matter, and a comparison set that includes only stars; and we compare the statistical properties of the two distinct sets. Our comparison demonstrates that the rates of change of the image fluxes are potentially a powerful diagnostic for the presence of planetary-mass lumps of dark matter in the lens galaxy. In \S\ref{section:ogledata} we analyse the OGLE data for \source\ and then compare their statistical properties to those of the simulations in \S\ref{section:analysis}, finding a preference for the simulations that include the planetary-mass lumps of dark matter. In order to be able to make that comparison it was necessary to first apply low-pass filtering to the OGLE data, and consequently we are unsure how strong our result really is. Section \S\ref{section:discussion} discusses how that situation can be improved, arguing for daily observations with a large, space-based telescope. Our summary and conclusions are presented in \S\ref{section:conclusions} .

\section{Macro-model of the \source\ lens}\label{section:macromodel}
\source\ is a well-studied quasar at a redshift of $z_s=1.695$, strongly lensed by a foreground spiral galaxy at $z_l=0.0394$ into a highly symmetric configuration of four images \citep[a ``quad'';][]{1985AJ.....90..691H}. The four images are all seen through the stellar bulge of the lens galaxy, and they display strong independent brightness fluctuations due to microlensing \citep{1989AJ.....98.1989I, 1991AJ....102...34C}, superposed on flux variations that are common to all images.

A detailed study of the lensing galaxy was undertaken by \cite{2010ApJ...719.1481V} who constructed axisymmetric dynamical models of the stellar bulge, constrained to match both the observed surface brightness profile and the kinematic profiles determined from integral field spectroscopy. Separately, \cite{2010ApJ...719.1481V} also modelled the total mass distribution in the lens that is required to reproduce the observed image positions of the quasar. That total mass distribution proved to be consistent with the stellar mass distribution, so that no dark matter is required to explain the macrolensing that is observed. Some dark matter may be present, but it is sub-dominant. Considering the uncertainties in the stellar mass-to-light ratio of the stars in the bulge, \cite{2010ApJ...719.1481V} concluded that the dark matter fraction in the lens galaxy (if assumed to be constant) is constrained to be $\la 20$\%\ of the total.

In this paper we adopt the isothermal lens model of \cite{2010ApJ...719.1481V}, for which the total convergence $\kappa_t$ -- {\it i.e.}, surface density in units of the critical surface density for gravitational lensing, for this lens-and-source combination -- of each image is as given in Table~\ref{table:macroparameters}. The appropriate shear $\gamma$ is equal to the total convergence, $\gamma=\kappa_t$ in each case.

\subsection{The dark matter contribution}\label{subsection:coreddarkhalo}
Models of galactic dynamics \citep[e.g.][]{2008gady.book.....B} treat the constituent stars as point masses, because even over times as long as the age of the universe physical collisions between stars are highly improbable. But that is not true of the dark matter lumps under consideration in this paper, which have a much lower column-density than stars. For example: our Sun has a mean column-density ${\rm M}_\odot/\pi {\rm R}_\odot^2\sim 10^{11}\;{\rm g\, cm^{-2}}$ whereas, as described in the Introduction, we are considering gas clouds whose peak column-density is  $\Sigma_0 \la 10^5\;{\rm g\, cm^{-2}}$. At low column-densities, physical collisions between gas clouds limit their possible contribution to any virialised dark matter halo \citep{1996ApJ...472...34G}.

That constraint was investigated by \cite{1999MNRAS.308..551W} who modelled the evolution of the density profile for a spherical dark matter halo comprised entirely of dense gas clouds (no non-baryonic dark matter) --- a scenario consistent with the approach taken in this paper.  Starting from a singular isothermal sphere, the evolved profile was found to be a cored isothermal sphere with the core radius depending on both the velocity dispersion of the halo and the mean column-density of the individual clouds. Of course material from the clouds which have undergone collisions simply becomes part of the pool of visible matter, so this model predicts a specific relationship between the circular speed of a dark halo and the total mass of visible material that should be present within it. Using published data, \cite{1999MNRAS.308..551W} demonstrated that the predicted, power-law relationship between visible mass and circular speed is obeyed by real galaxies, and that this provides a simple physical basis for the well-known Tully-Fisher relation for star-forming galaxies.\footnote{It is now widely accepted that a relationship between visible mass and halo circular speed does indeed underlie the Tully-Fisher relation; it has since become known as the ``Baryonic Tully Fisher Relation'' \citep[following][]{2000ApJ...533L..99M}.}

\begin{table}
	\centering
	\caption{Total convergence values used for each of the four images of \source. In our stars+``planets'' microlensing simulations, the optical depth in $10^{-4}\,\mathrm{M}_\odot$ point mass lenses was set to $\kappa_p=0.043$ for each of the images, with the remainder, $\kappa_t-\kappa_p$, comprised of $0.1\,\mathrm{M}_\odot$ stars. For our ``stars only'' microlensing simulations we set $\kappa_p=0$.}
	\label{table:macroparameters}
	\begin{tabular}{lcccc} 
		\hline
		Image & A & B & C & D \\
		Total $\kappa=\gamma$ & 0.396 & 0.391 & 0.715 & 0.604 \\
		\hline
	\end{tabular}
\end{table}

The success of such a simple physical model gives support to the idea that dark halos are indeed made up of dense gas clouds. But for the purposes of the present paper the key point is that by matching to data \cite{1999MNRAS.308..551W} was able to pin down the mean column-density of the individual gas clouds: $\langle\Sigma\rangle\equiv M/\pi R^2 \simeq 140\;{\rm g\, cm^{-2}}$. And consequently the core radii of these collisional dark matter halos can be predicted from their circular speeds alone. For the galaxy that is responsible for lensing \source, the isothermal model of \cite{2010ApJ...719.1481V} has a velocity dispersion of $170\;{\rm km\,s^{-1}}$, corresponding to a dark halo core radius of approximately $7.1\;{\rm kpc}$ \citep{1999MNRAS.308..551W}. That radius is several times larger than the projected separation of each of the four images from the centre of the lens galaxy, so the expected mean column-density of the dark halo is approximately the same for all of them and approximately equal to the column at zero impact parameter. It is straightforward to integrate the density of a cored isothermal sphere along the line of sight, and expressing the result in terms of the critical surface density gives a central convergence (optical-depth) of $\kappa_p\simeq 0.043$ in planetary-mass gas clouds.

The dark column just quoted is between $6$\%\ and $11$\%\ of the total columns given in Table \ref{table:macroparameters} for the four images of \source. Although \cite{2010ApJ...719.1481V} did not formulate constraints for the particular case of a cored dark halo, their estimate of $\la 20$\%\ for a {\it constant\/} dark matter fraction is well above all our $\kappa_p/\kappa_t$ values, so it is likely that our model could yield an acceptable fit to their data. In fact the core radius of our model is sufficiently large that the dark halo surface density is almost constant in the region where the images of \source\ are formed, and consequently any lensing constraints are subject to the notorious mass-sheet degeneracy \citep[see][for example]{2013A&A...559A..37S}. We also note that in our model the mean microlens mass is given by $\langle m_{lens}\rangle\simeq (1-\kappa_p/\kappa_t)\langle m_*\rangle$, which is 89-94\% of the mean mass of the stars in the bulge, $\langle m_*\rangle$. In other words: the mean microlens mass in our model is just that of a low-mass star, consistent with the findings of \cite{2000MNRAS.315...51W}.

\section{Microlensing simulations}
\label{section:simulations}
We generated two sets of magnification maps as a function of the source position using a bespoke ray tracing code, with the first putting all local mass density into stars only and the second allowing about ten per cent of it to be in the substellar component. It is computationally challenging to use a large dynamic range in the mass of individual lenses and for this reason we used the top of the range of cloud masses preferred by \cite{2022MNRAS.513.2491T}, $M_p=10^{-4}\,\mathrm{M}_\odot$ as the ``planet" mass, and the mass of each of the stellar lenses was fixed at $0.1\,\mathrm{M}_\odot$; Appendix~\ref{appendix:sizenmass} discusses the effect of adopting a lower mass for the ``planets''. Table~\ref{table:macroparameters} lists the lensing convergence (equal to the shear, for our isothermal macro model) for the four images of the quasar. For the stars+``planets" set, a value of $\kappa_p=0.043$, the same for all images because of their highly symmetric configuration, is comprised of ``planets" and the rest is put into stars; for the star-only set, all of the convergence is in stars.

\begin{figure*}
\centering
\includegraphics[width=160mm]{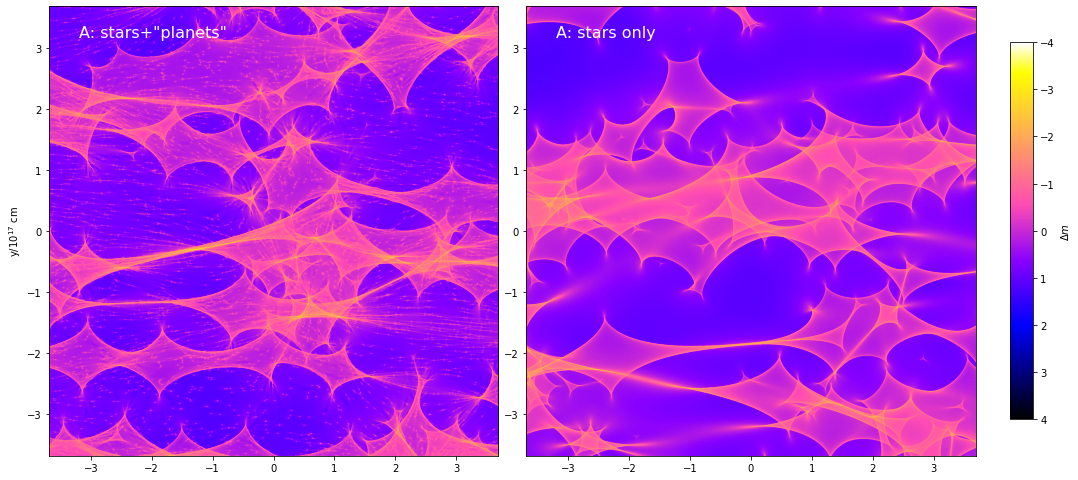}
\includegraphics[width=160mm]{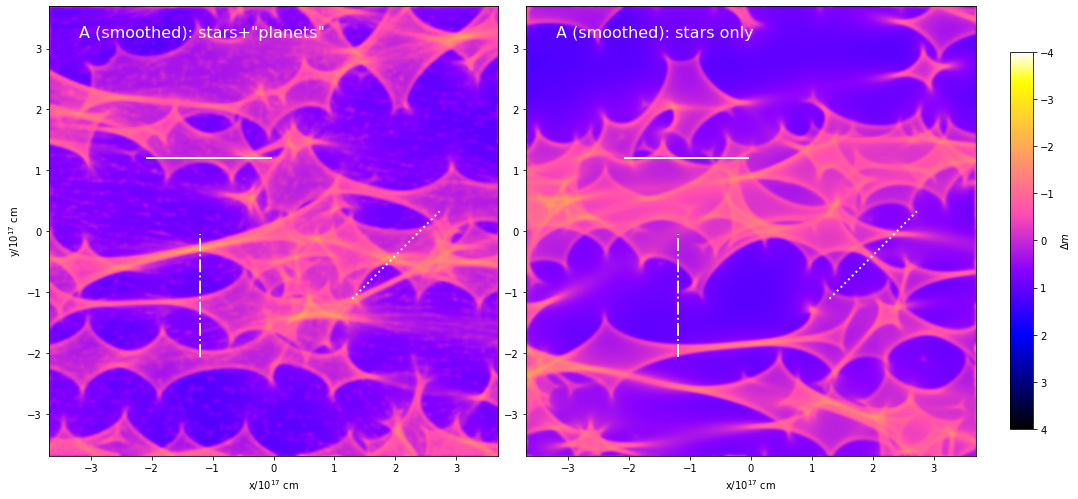}
\includegraphics[width=150mm]{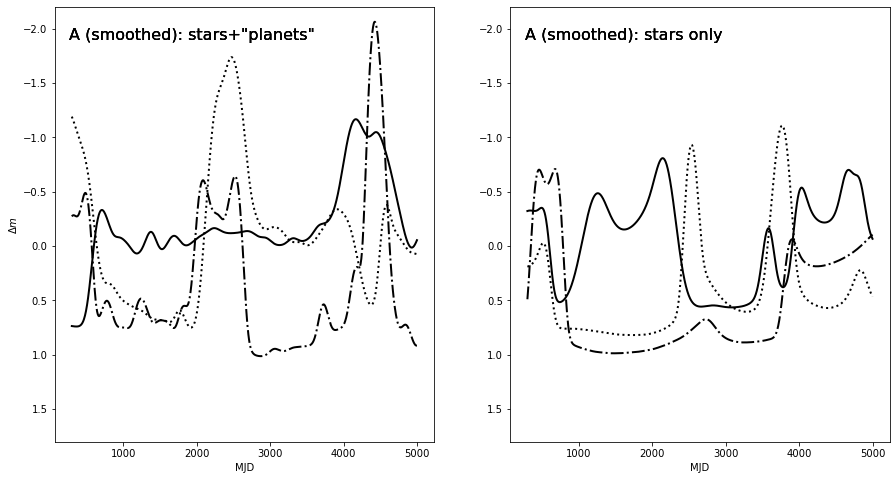}
\caption{Microlensing magnification, in magnitudes, as a function of the source position for the positive parity image A. The right column assumes that all mass is in stellar-mass objects while the left includes a contribution from planetary-mass (nano-) lenses. The top row is for a point (or, more precisely, pixel-sized) source while the maps in the middle are convolved with a Gaussian disc of half-light radius $R_s=3\times10^{15}\,\mathrm{cm}$. The bottom row provides examples of light curves for a source moving with an effective transverse velocity of $5\times10^3\,\mathrm{km}\,\mathrm{s}^{-1}$ along (solid curve), across (dot-dashed) and at 45 degrees (dotted) to the large-scale shear stretching direction; the corresponding tracks are also marked in the middle row.}
    \label{figure:magmapsA}
\end{figure*}

\begin{figure*}
\centering
\includegraphics[width=160mm]{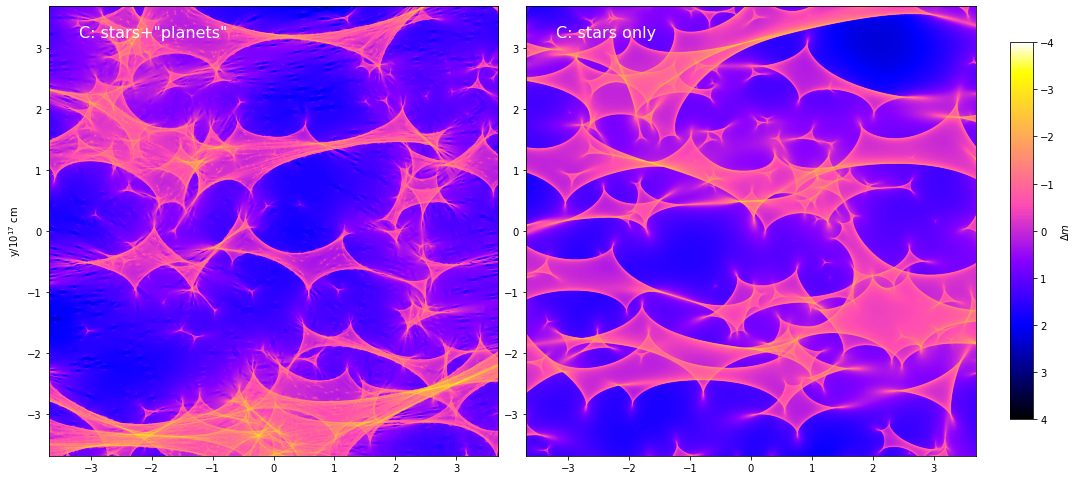}
\includegraphics[width=160mm]{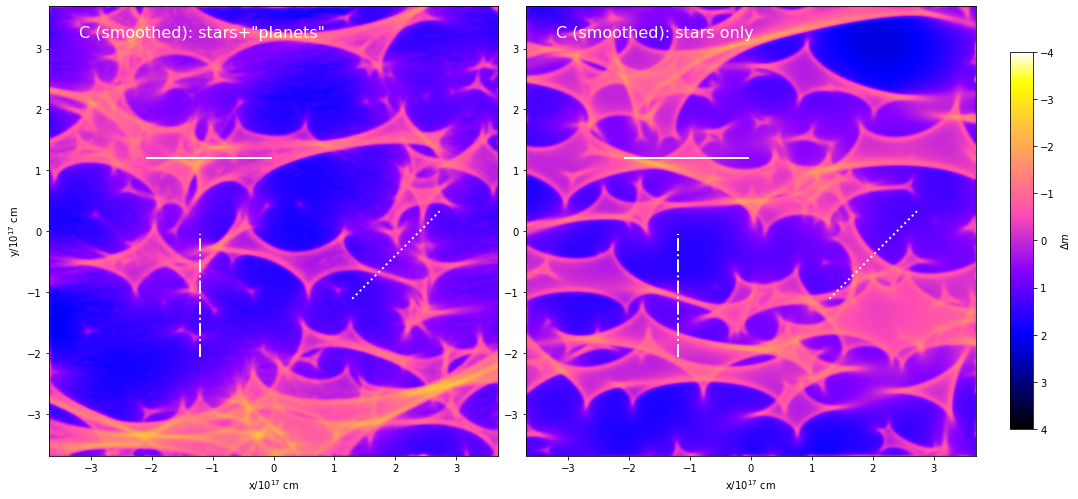}
\includegraphics[width=150mm]{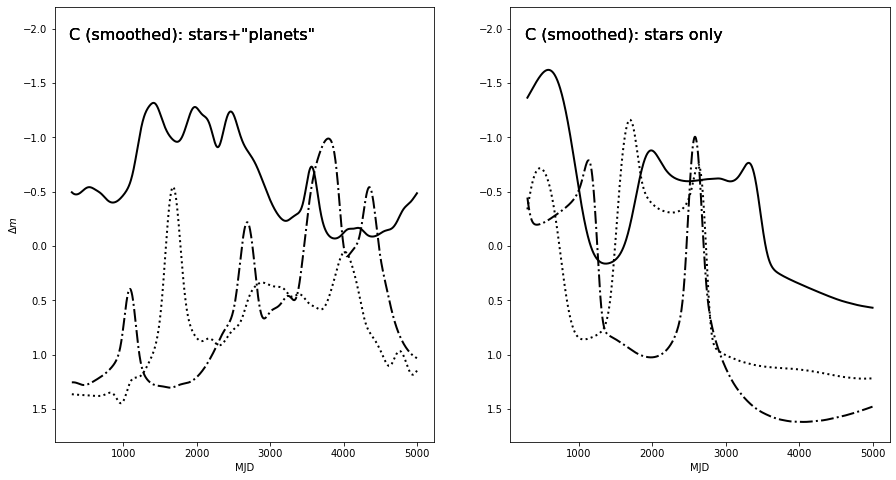}
\caption{Same as Figure~\ref{figure:magmapsA} for image C, of negative parity.}
    \label{figure:magmapsC}
\end{figure*}

The resulting magnification maps are $2048\times2048$-pixel squares, covering $7.4\times10^{17}\,\mathrm{cm}$ (four Solar-mass Einstein radii, projected onto the source plane) on a side. This results in a resolution of $\sim160$ and $\sim 5$ pixels per stellar and ``planetary'' Einstein radius, respectively. The results are displayed in Figure~\ref{figure:magmapsA} and~\ref{figure:magmapsC} for one positive and one negative parity image in the system  -- A and C, respectively; similar figures,~\ref{figure:magmapsB} and~\ref{figure:magmapsD}, for images B and D are included in the Appendix~\ref{appendix:bdmaps}. The maps are then convolved with the intensity distribution over the source profile, chosen as a disc with a Gaussian brightness profile of half-light radius $R_S=3\times10^{15}\,\mathrm{cm}$, or approximately eight pixels.

As discussed in Appendix~A of \citet[][see, particularly, their Figure A1]{2022MNRAS.513.2491T}, the information that is currently available in the literature favours quasar optical half-light radii in the range $(1-5)\times 10^{15}\;{\rm cm}$. Our chosen size for \source\ lies towards the upper end of that range, reflecting the fact that it is one of the more luminous quasars known. The photometric estimate for \source\ half-light radius along the lines of their Appendix~A returns values between $2.2$ and $4.8\times10^{15}\,\mathrm{cm}$, depending on the image chosen, and $3\times10^{15}\,\mathrm{cm}$ is a representative estimate. We discuss the effect of changing the assumed source size in Appendix~\ref{appendix:sizenmass}.

In order to convert magnification gradients into temporal derivatives in the image fluxes, we also need an estimate of the velocity at which the source moves relative to the magnification pattern. As discussed by \citet{2004ApJ...605...58K}, the observer contribution to the effective transverse velocity is inferred to be small, because \source\ lies not far from the axis of the observed dipole in the cosmic microwave background; and because the lens is at much lower redshift than the source the source contribution is also expected to be unimportant. With a distance ratio (source:lens) of approximately 11, a plausible value of around $\sqrt{2}\times 300\;{\rm km\,s^{-1}}$ for the transverse velocity of the lens galaxy leads us to our adopted value for the source-plane projected transverse velocity of $5{,}000\;{\rm km\,s^{-1}}$.

\section{Comparison data and filtering}
\label{section:ogledata}

The Optical Gravitational Lensing Experiment (OGLE) has been monitoring \source\, since 1997, resulting in a dense and homogeneous set of accurate photometric measurements for all four images of the system, spanning over a decade in time \citep{2006AcA....56..293U}. These data are available to the astronomical community in the OGLE Internet data archive\footnote{http://ftp.astrouw.edu.pl/ogle/ogle2/huchra/, http://ftp.astrouw.edu.pl/ogle/ogle3/huchra/}, where we downloaded it from. The data contain magnitude measurements for the four images and their uncertainty estimates with, on average, one photometric point every few days, with some seasonal gaps. The photometric uncertainty ranges from under 10 millimagnitudes in the brightest image, A, to about 30 millimagnitudes for image D, which is 1-2 magnitudes fainter. 

\begin{figure*}
\centering
\includegraphics[width=180mm]{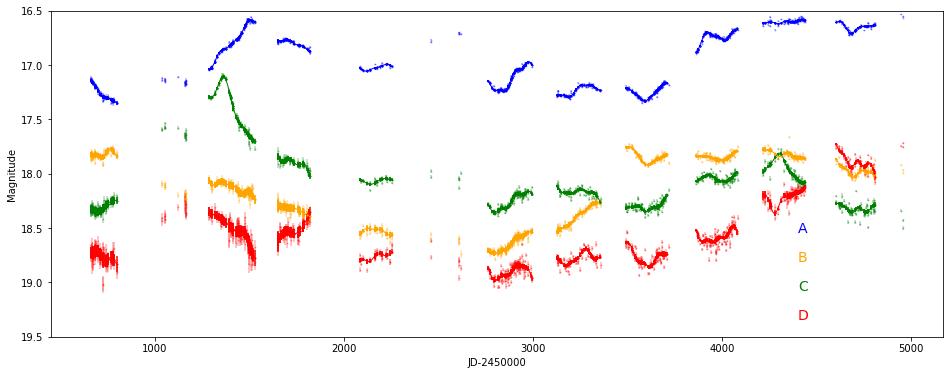}
\includegraphics[width=180mm]{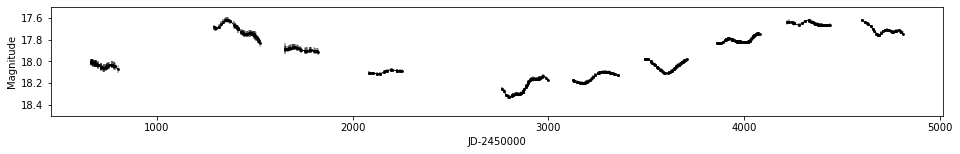}
\includegraphics[width=180mm]{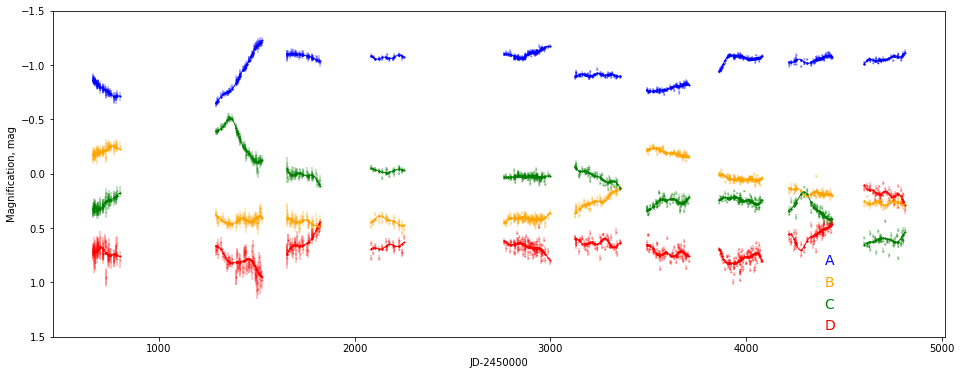}
\caption{The top panel displays OGLE light curves for the images of \source, shown with the original error bar estimates along with their smoothed versions, obtained by fitting harmonic series with periods above 60 days. The middle panel presents the mean of the four (smoothed) light curves, which we use as an estimate of the intrinsic variations in the source. The bottom panel shows the difference between the data at the top and middle panels, interpreted in this paper as an estimate of the microlensing magnification. We dropped seasons with only a handful of measurements (near MJD51100 and MJD52600) where low-pass filtering makes little sense.}
    \label{figure:lightcurves}
\end{figure*}

These OGLE data, which are displayed in the top panel of Figure~\ref{figure:lightcurves}, form our starting point for comparing the statistics of observed light curve derivatives with the results of our microlensing simulations. There are, however, two issues that we need to face up to when making use of these data. One is the intrinsic variability of the source, which contributes to the observed light-curve derivatives for each of the images, and the other is photometric errors, which make estimates of the magnification derivatives noisy.

The latter is important for us because even the best of the raw photometric sequences (image A) manifests photometric noise at the level of $\sim 10\;{\rm mmag}$ differences over spacings $\sim 3\;{\rm days}$ --- corresponding to gradient noise $\sim 3\;{\rm mmag\,day^{-1}}$ and, as we show later (\S5, and in particular Figure~\ref{figure:magstats}), the predicted signals also manifest at the level of $\sim 3\;{\rm mmag\,day^{-1}}$. To deal with the noise we low-pass filter the individual light curves, fitting them with harmonic time series with periods limited from below. We chose a limit of 60 days, recognising the scale of the structure we want to remain sensitive to and the data cadence, but the results are fairly insensitive to the smoothing scale choice.

To estimate the intrinsic variations, we simply form the average of the four low-pass filtered light curves. (Note that the time delays between the images are expected, and tentatively measured, to be small --- of order hours, see \citealt{1998MNRAS.295..488S, 2003ApJ...589..100D, 2006A&A...447..905V, 2014MNRAS.442..428W}). The result, which is shown in the middle panel of Figure~\ref{figure:lightcurves}, is inevitably contaminated by  microlensing in the sense that it includes, at each epoch, the microlensing magnification averaged over all four images.

Subtracting our estimate of the intrinsic variations from the light-curves of the individual images then provides us with our estimates of the microlensing signals for each of the four images separately, as shown in the lower panel of Figure~\ref{figure:lightcurves}. These individual microlensing sequences are (to the extent that relative time-delays are negligible) free of contamination by the intrinsic variations of the quasar. However, because our estimate of the intrinsic variations includes the mean of the microlensing magnifications, the individual microlensing sequences are themselves also contaminated by the (negative of) the mean microlensing magnification. This framework for data analysis is new, so in Appendix~\ref{appendix:differences} we repeat the statistical analysis of Section~\ref{section:analysis} using the more traditional approach \citep[e.g.][]{2023ApJ...954..172E} involving magnitude differences between image pairs. That path leads to essentially identical conclusions.

\section{Light curve statistics}
\label{section:analysis}

Figures~\ref{figure:magmapsA} and~\ref{figure:magmapsC} compare the microlensing magnification maps in the stars-only and stars+``planets" cases. Despite ``planets'' only making about $\sim10$ per cent contribution to the optical depth, the effect of adding them is striking. The extended low-magnification ``valleys" and high-magnification ``plateaus" with a characteristic size of $10^{16-17}\,\mathrm{cm}$, corresponding to years-long time spans for realistic effective velocities, develop small-scale bumps; and, in the valleys for negative parity images, dips of a characteristic shape. At a ``planet" mass of $10^{-4}\,\mathrm{M}_\odot$ and parameters appropriate for \source, their size is $\sim10^{15-16}\,\mathrm{cm}$, right at the scale that is big enough to not be completely smoothed out by the finite source size effects, yet small enough to be probed with observations on $\sim\,$weekly timescales.

\begin{figure}
\centering
\includegraphics[width=\columnwidth]{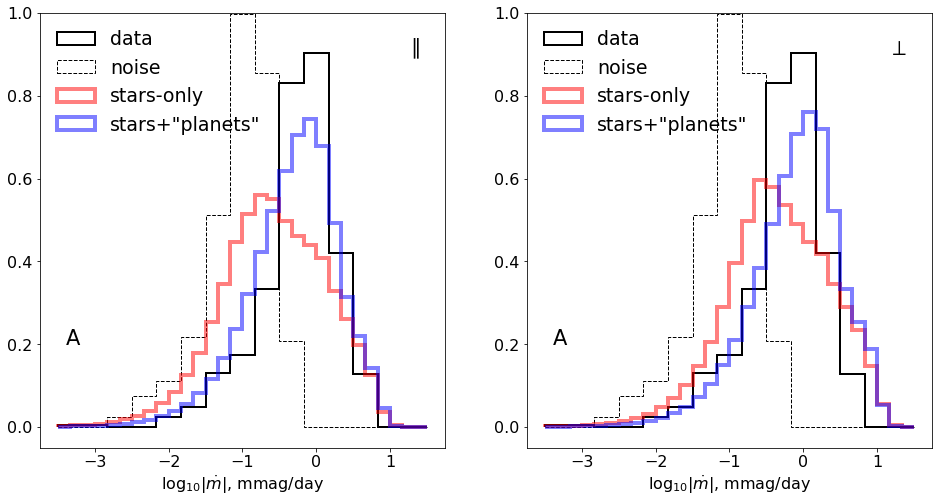}
\includegraphics[width=\columnwidth]{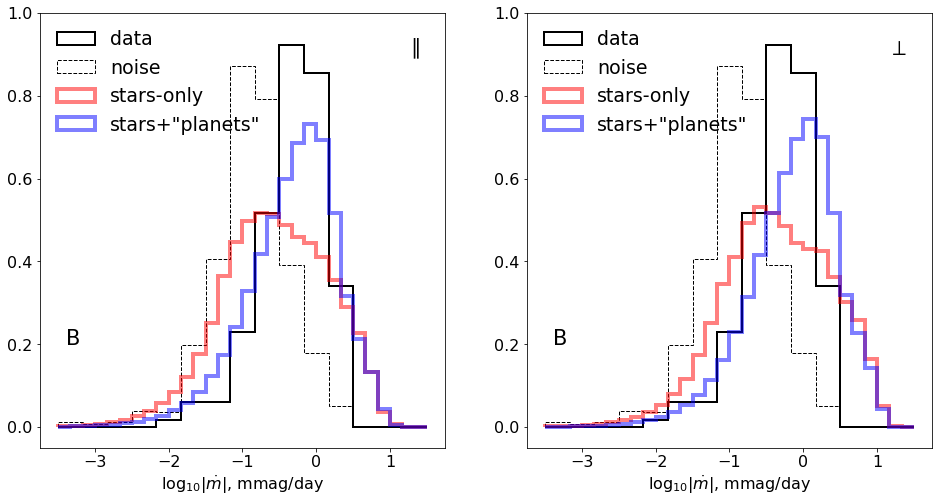}
\includegraphics[width=\columnwidth]{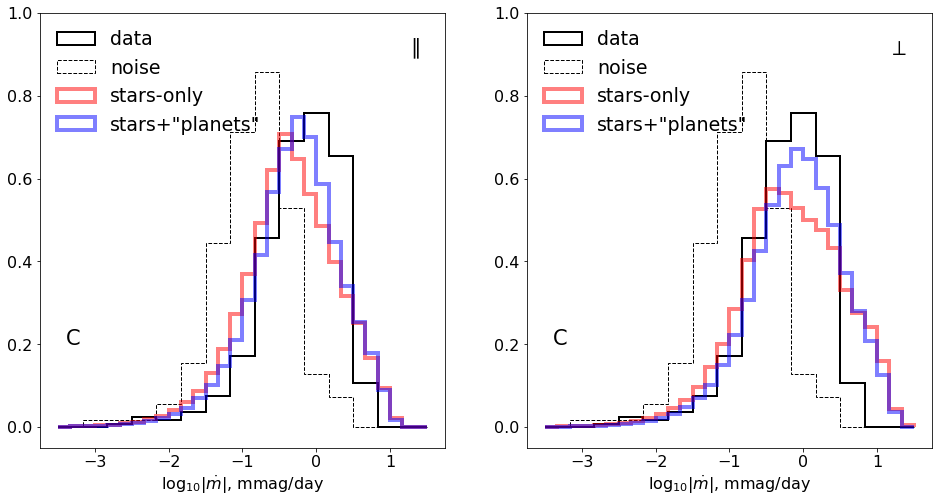}
\includegraphics[width=\columnwidth]{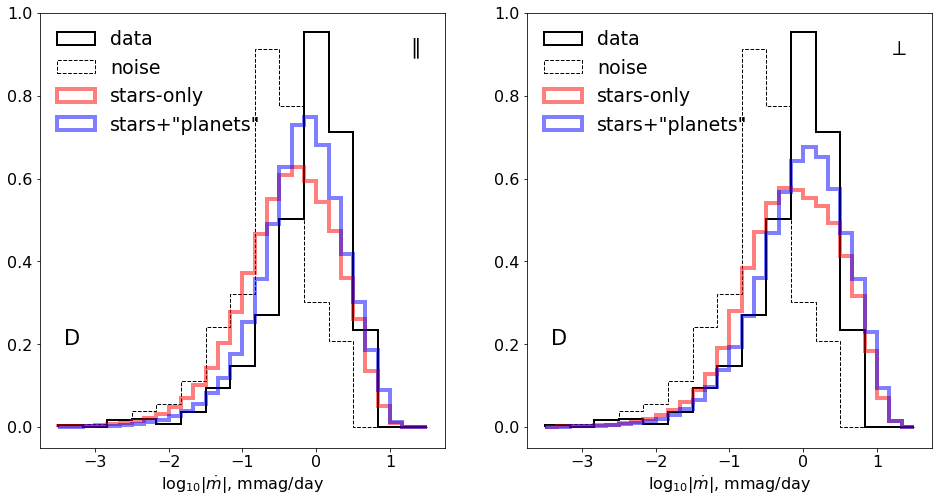}
\caption{Histograms showing the distribution of the two components of the gradient of the (smoothed) magnification maps generated for each macro-image: A, B, C and D, from top to bottom. The left-hand column shows the component parallel ($\parallel$) to the shear, and the right-hand column shows the perpendicular ($\perp$) component. Blue lines show the case with both stars and planetary-mass lenses; red lines show the case with stars only. In both cases the histograms are averages over the whole map. A source velocity of $5\times10^3\,\mathrm{km}\,\mathrm{s}^{-1}$ is used to convert the gradient into time derivatives. The solid black line shows the observed distribution of light curve derivative in smoothed OGLE data from Figure~\ref{figure:lightcurves} and the dashed line is for simulated light curves containing noise at the level reported by OGLE -- processed in the same way as the actual data (note that the histograms are normalised).}
    \label{figure:magstats}
\end{figure}

We found the magnification distributions of the stars+``planets" and stars-only cases to be broadly similar. This is consistent with the expectation \citep{1992ApJ...386...19W, 1995MNRAS.276..103L, 2001MNRAS.320...21W} for the magnification distribution to be only weakly sensitive to the microlens mass distribution (apart from the moderating effects of a finite source size). However, we note that \cite{2004ApJ...613...77S} demonstrated the situation to be significantly more complex, and the issue has not been resolved yet. It is clear though that the small-scale structure significantly affects the statistics of magnification derivatives, as can be seen in Figure~\ref{figure:magstats}. This is particularly the case for the positive-parity images A and B, whose stars-only magnification maps contain extensive low-magnification valleys with very little small-scale structure --- especially where source tracks happen to align with the direction along which the micro-caustic structure is stretched by the large-scale shear. The small-scale ``planetary" structure disturbs such quiet regions, in the process making the derivative statistics somewhat less anisotropic. Derivatives both along and across the large-scale stretching direction are enhanced, though the contrast is greater for the former.

The large effect of ``planets" on the distribution of magnitude derivatives suggests using that statistic to assess which of the two scenarios matches best to the OGLE data. In practice, we can only measure the derivative along the particular direction in which the image tracks through the magnification map -- {\it i.e.}, a linear combination of the two components of the gradient. The orientations of the actual tracks in the images of \source,  and thus the relative contributions of the two components, are not well constrained \citep{2010ApJ...712..658P, 1999MNRAS.309..261W, 2004MNRAS.352..125T}; and the random motion of the microlenses in the bulge of the lens galaxy is also expected to contribute significantly to the time derivative \citep{2000MNRAS.312..843W}. However, the change in shape of the histograms that is effected by the introduction of ``planets'' is qualitatively similar for both components of the magnification gradient, which encourages a comparison with the data even in the absence of a unique value for the transverse velocity. Intriguingly, the observed distributions, shown in Figure~\ref{figure:magstats} as black curves do look more like the blue, stars+``planets" and less like the red, stars-only histograms. Bearing in mind that only the horizontal position and not the shape of the simulated histogram depends on the unknown velocity, the observed narrower distributions and their sense of asymmetry appear consistent with the presence of ``planets", particularly for the A and B images, in which the difference between the two cases is most pronounced.

\begin{figure}
\centering
\includegraphics[width=42mm]{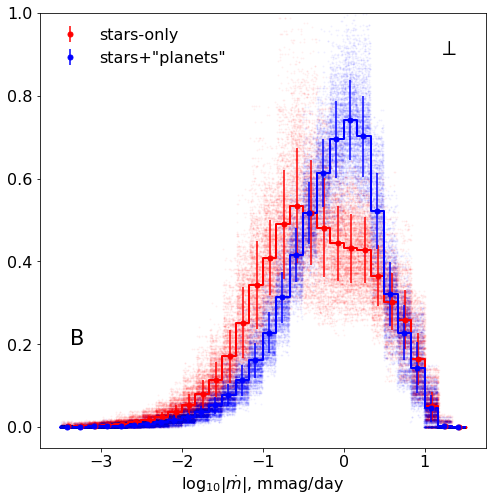}
\includegraphics[width=42mm]{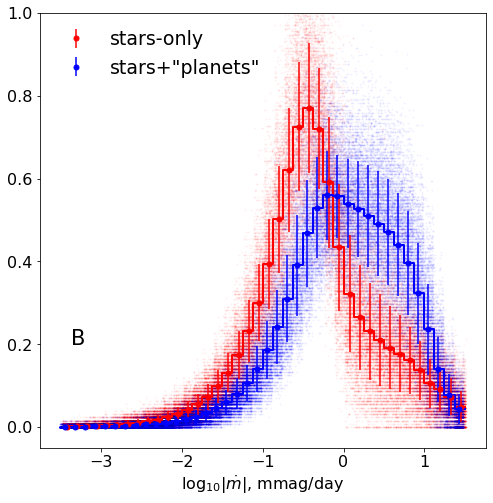}
\includegraphics[width=42mm]{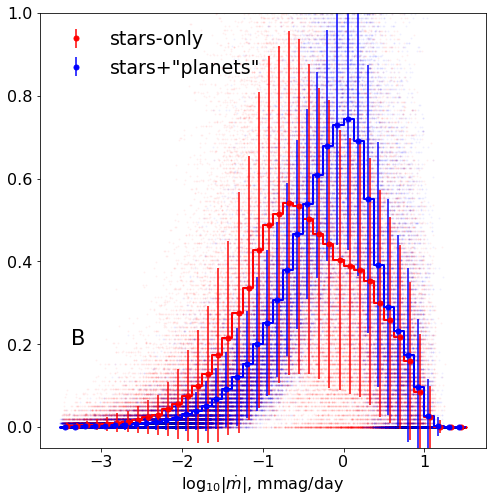}
\includegraphics[width=42mm]{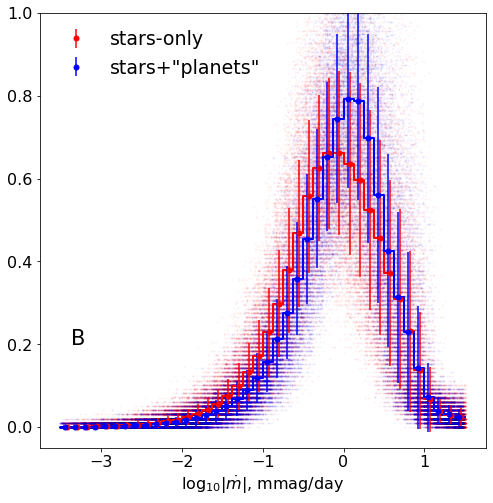}
\caption{Degradation of the attribution capacity of the magnification derivative histograms with practical constraints. The top left panel shows the average (as a step line plot) and standard variation (as error bars) of the magnification distributions among a set of 2048 light curves covering the full magnification maps along its columns; scattered points, shifted randomly within each bin for visual clarity, represent individual histograms. The top right panel accounts for the unknown orientation of tracks and the bottom left panel mimics the sampling pattern of the actual OGLE data set. The bottom right panel displays the effect of adding to the simulated magnification of a Gaussian noise with the amplitude reported by OGLE as their photometry estimated accuracy; for noisy data, the derivatives are computed after low-pass filtering them in the same way as done for the actual OGLE light curves. Note the effect of noise on the average of the statistic.}
    \label{figure:degradation}
\end{figure}

\begin{figure}
\centering
\includegraphics[width=\columnwidth]{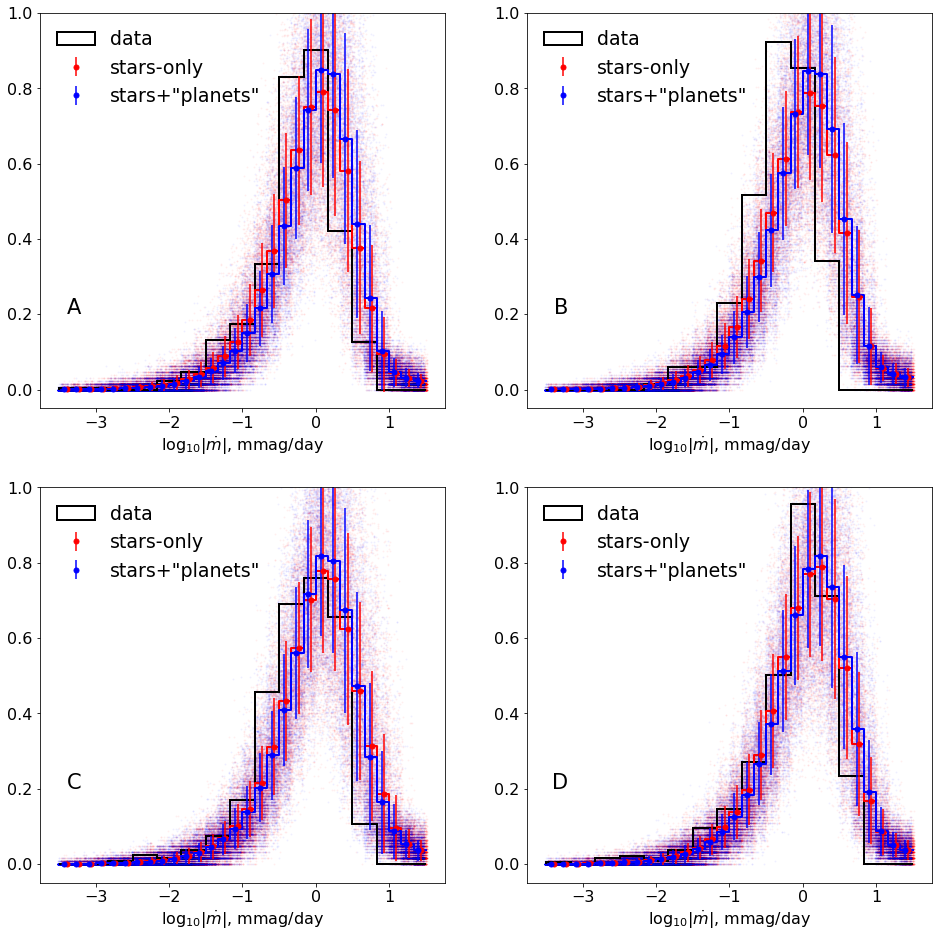}
\caption{Predicted distribution, and its variation, for the magnification derivative in the low-pass filtered simulated noisy light curves from which an estimate of the intrinsic variation in each set of four light curves has been subtracted. The black lines show the derivative distribution of the actual light curves processed in the same way.}
    \label{figure:datahistograms}
\end{figure}

However, at present the preference for the stars+``planets" case is not much more than suggestive. Although the predicted derivative distributions for our entire simulated magnification maps are clearly distinct, there is substantial variation in these distributions among the various possible realisations of individual tracks within the maps. This is illustrated in the upper left panel of Figure~\ref{figure:degradation}, which plots the vertical derivative distributions for 2048 individual columns of the maps for image B, shown in the middle panels of Figure~\ref{figure:magmapsB} -- a case that, from Figure~\ref{figure:magstats}, appears most favourable for distinguishing the two cases. Even for these light curves, which are ideally matched to the extent and resolution of the magnification maps, the scatter is so large that it seems challenging to attribute any particular distribution to either red or blue set. Subsequent panels of the figure add, cumulatively, the effects of the unknown track orientation (upper right), OGLE-like sampling (bottom left) and OGLE-like noise (bottom right -- these light curves are low-pass filtered before the derivatives are estimated). Finally, Figure~\ref{figure:datahistograms} compounds the above effects by forming and subtracting the estimate of the intrinsic light curve (inevitably contaminated by the microlensing) for each simulated set of four light curves, and computes their derivative distribution; the statistics of the resulting histograms are compared to the observed derivative distribution for all images.

The progressive degradation seen in the panels of Figure~\ref{figure:degradation} suggests that it is mostly the noise that deprives the magnification derivatives of their power to discriminate between the stars-only and stars+``planets" cases. Moreover, unlike poor sampling, which only increases the expected variation in the statistic without affecting the mean value much, the addition of (as assumed here, white) noise distorts the latter, which diminishes prospects to improve the discriminating power with better sampling. This suggests that it might not be possible to rule out (or in) a subdominant low-mass lens population using only the available data. On the flip side, reducing the measurement noise would be on the top of the wish list for a future data set that would allow a definite test for the presence of nanolenses in \source, followed by improved sampling. Encouragingly, this is also the conclusion of a study (G.F.~Lewis et al, in preparation) using machine learning algorithms and human classifiers to distinguish between simulated stars-only and stars+``planets'' cases. Relying on light curve features that are not fully captured in the statistics that the present paper focuses on, the algorithms are able to outperform the human classifiers in terms of false (``planet'') positives if the noise is sufficiently reduced. 

\section{Discussion}\label{section:discussion}

\subsection{X-ray line shifts in lensed quasars}
A substantial population of free-floating planetary-mass objects has previously been suggested to exist in external galaxies \citep{2018ApJ...853L..27D,2019ApJ...885...77B}, based on time-varying shifts in the Fe~K$\alpha$ line energy that were observed for several gravitationally lensed quasars \citep{2017ApJ...837...26C}. Those studies are qualitatively consistent with the picture advanced in the present work, but quantitatively they are inconsistent because \citet{2018ApJ...853L..27D} and \citet{2019ApJ...885...77B} estimated the population to be much smaller, and dynamically unimportant. However, those estimates rely on a specific model for the quasar's X-ray line emission -- intensity as a function of both spatial and velocity coordinates -- and at present our knowledge of that intensity structure is not reliable.

\subsection{Other multiply imaged quasars}
In this paper we have concentrated our attention on just one multiply-imaged quasar, \source. But many strongly lensed quasars are known \citep[e.g.][]{2010ARA&A..48...87T, 2018MNRAS.475.2086A, 2023MNRAS.520.3305L}, and dozens of those now have high quality light-curves available, which raises the question of whether there are any other systems that might be useful in searching for a planetary-mass population in the lens galaxy? Ultimately that question must be answered on a case-by-case basis; however, the general outlook can be readily summarised: \source\ is particularly well suited to our purposes because of the exceptionally low redshift of the lens galaxy. There are two main considerations. First, for planetary-mass lenses the predicted amplitude of the photometric signal is quite modest even for \source, and for more distant lens galaxies the amplitude decreases because the size of the magnification structure goes down relative to the size of the source. Secondly, the very low redshift of the lens in \source\ means that its contribution to the (source plane) effective velocity is expected to be very large -- as discussed by \citet{2004ApJ...605...58K} -- and the nanolensing event rate is expected to be correspondingly high. Thus for \source\ the nanolensing events are expected to be bigger and more frequent than for a typical lensed quasar. Moreover these differences are substantial, because a typical lens -- e.g. in the sample studied in the COSMOGRAIL project \citep{2020A&A...640A.105M} -- has a redshift that is an order of magnitude greater than in \source.

All this is not to say that \source\ is the best possible system for this type of study, as there's no fundamental barrier to the existence of a quasar that is lensed by an even closer galaxy (and with a greater transverse velocity, and a smaller source, etc.). New, deeper surveys for lensed quasars have the potential to discover a system that is superior to \source\ in these respects.

\subsection{A call for new data on \source}
\label{subsection:callfornewdata}
The influence of planetary-mass lenses can be clearly seen in our simulations -- in the magnification maps, the simulated lightcurves, and the histograms of magnification gradients -- but that is in part because the calculations are very precise. In fact each nano-lens produces only a small amplitude photometric signal, comparable to the error bars on the OGLE data --- hence the need for smoothing, and motivating our application of low-pass filtering to the data. Although low-pass filtering is a legitimate procedure to use, it does create some concerns. First, it means that the models are being compared with a signal that is one step removed from the measurements themselves. Secondly, it is possible that the  smoothed data might be limited by low-level systematic errors -- associated with, for example, lunation or variations in atmospheric transparency -- that can masquerade as a microlensing signal. Consequently we cannot gauge the level of support for the model of dark matter in the form of planetary-mass gas clouds given by the currently available OGLE data.

The problems just mentioned can be addressed by obtaining a new dataset. With a large telescope, and exposure times of only a few minutes, the photon noise limit can be better than $1\;{\rm mmag}$, even for the faintest of the four macro-images of \source. And because the requisite exposure times are short such observations can be undertaken with a high cadence (e.g. daily) for the whole observing season. In that way one can ensure that if the predicted nanolensing fluctuations are manifest in the lightcurves then they would be well sampled by the observations and would be measured with high signal-to-noise. 

Although high signal-to-noise and high cadence are both important they are not sufficient to guarantee a clear detection of the predicted signal (or tight limits on the absence thereof) in \source: it's also essential to ensure that systematic errors -- in particular, fluctuating systematic errors -- are reduced to a level where they are insignificant. That is difficult to achieve at high levels of precision because of, for example, low-level residual variations in the sky background, the atmospheric transparency and the point-spread function. However, by far the largest contributions to those effects come from the Earth's atmosphere, and if we observe from space it is possible to keep systematic errors to a level well below $1\;{\rm mmag}$. In the near future the large aperture space telescopes that will permit targetted observations include {\it JWST\/} \citep{2023arXiv230101779M} and the {\it Nancy Grace Roman\/} space telescope (WFIRST) \citep{2019arXiv190205569A}; but both of these are primarily infrared telescopes, which is not ideal for quasar microlensing studies because the size of the emission region increases with wavelength. The {\it Hubble Space Telescope\/} \citep{2020RAA....20...44W}, on the other hand, has a large aperture and can observe in the visual, blue or even ultraviolet bands --- it appears to be the best of the available space telescopes for this task.

\section{Summary and Conclusions}
\label{section:conclusions}

Planetary-mass gas clouds are able to explain the statistically symmetric and largely achromatic variability of field quasars if they make a significant contribution to the dark matter budget, while nevertheless evading Galactic constraints on compact low-mass lenses \citep{2022MNRAS.513.2491T}. A population of such clouds would form cored \emph{baryonic} dark matter profiles in the galactic context and, at the characteristic cloud column density implied by the observed properties of star-forming galaxies \citep{1999MNRAS.308..551W}, the clouds remain strong lenses when placed outside of the Local Group, at $z\gtrsim 0.01$. 

In this paper, we considered the effect of such planetary-mass nanolenses on the light curves of \source\ -- one of the best-studied, multiply-imaged quasars \citep{1985AJ.....90..691H}. The four images of this quasar are seen through the bulge of a much nearer lens galaxy and display prominent microlensing signals \citep{1989AJ.....98.1989I}, which can be separated from any intrinsic variation in the background quasar source because the time delays between images are all very small. The macro-lensing model is tightly constrained by the observed image astrometry, surface photometry and integral field spectrometry of the lens galaxy \citep{2010ApJ...719.1481V}. And a long, homogeneous photometric record of the four images has been accumulated by the OGLE project \citep{2006AcA....56..293U}.

The nanolensing population is predicted to remain a subdominant mass component, providing about ten per cent of the total lensing convergence, the rest contributed by the stellar population of the bulge. We simulated magnification maps appropriate to the projected location of the four images in the bulge and compared the predicted light curves to the available OGLE data. The nanolenses do not change the distribution of the microlensing magnification in any dramatic way, but the magnification derivatives are strongly affected. At face value, the observed distributions in Figure~\ref{figure:magstats} do look more like the simulations containing nanolenses than those with stellar mass microlenses only. However, we do not believe this evidence to be conclusive for two main reasons. One is that the variation in the predicted derivative distribution from one realisation to another is substantial due to the limited time sampling of the available data. The second is the inevitable noise in the available photometry, which both necessitates low-pass filtering of the data, to infer magnification derivatives, and biases the model predictions ({\it cf.}~Figure~\ref{figure:degradation}). We therefore conclude that, although the OGLE data are consistent with the presence of a significant population of nanolenses in the bulge of the \source\ lens galaxy, we cannot categorically decide the matter at present.

The photometric noise thus emerges as a primary observational challenge to the attempted test and we call for a highly accurate set of light curves from a larger, preferably space-based telescope to resolve the issue. The images are sufficiently bright for the $\mathcal{O}(\mathrm{mmag})$ accuracy to be achieved with reasonable demand on the exposure time. We are also exploring alternative data characterisation methods designed to better handle the information not captured by the magnitude derivative statistics, such as machine learning classification.

\section*{Data availability}
The data underlying this article will be shared on reasonable request to the corresponding author.

\section*{Acknowledgements}
GFL received no financial support for this research. We thank the referee for thoughtful suggestions.

\bibliographystyle{mnras}
\bibliography{2237nano}

\begin{thebibliography}{}
\makeatletter
\relax
\def\mn@urlcharsother{\let\do\@makeother \do\$\do\&\do\#\do\^\do\_\do\%\do\~}
\def\mn@doi{\begingroup\mn@urlcharsother \@ifnextchar [ {\mn@doi@} {\mn@doi@[]}}
\def\mn@doi@[#1]#2{\def\@tempa{#1}\ifx\@tempa\@empty \href {http://dx.doi.org/#2} {doi:#2}\else \href {http://dx.doi.org/#2} {#1}\fi \endgroup}
\def\mn@eprint#1#2{\mn@eprint@#1:#2::\@nil}
\def\mn@eprint@arXiv#1{\href {http://arxiv.org/abs/#1} {{\tt arXiv:#1}}}
\def\mn@eprint@dblp#1{\href {http://dblp.uni-trier.de/rec/bibtex/#1.xml} {dblp:#1}}
\def\mn@eprint@#1:#2:#3:#4\@nil{\def\@tempa {#1}\def\@tempb {#2}\def\@tempc {#3}\ifx \@tempc \@empty \let \@tempc \@tempb \let \@tempb \@tempa \fi \ifx \@tempb \@empty \def\@tempb {arXiv}\fi \@ifundefined {mn@eprint@\@tempb}{\@tempb:\@tempc}{\expandafter \expandafter \csname mn@eprint@\@tempb\endcsname \expandafter{\@tempc}}}

\bibitem[\protect\citeauthoryear{{Agnello} et~al.,}{{Agnello} et~al.}{2018}]{2018MNRAS.475.2086A}
{Agnello} A.,  et~al., 2018, \mn@doi [\mnras] {10.1093/mnras/stx3226}, \href {https://ui.adsabs.harvard.edu/abs/2018MNRAS.475.2086A} {475, 2086}

\bibitem[\protect\citeauthoryear{{Akeson} et~al.,}{{Akeson} et~al.}{2019}]{2019arXiv190205569A}
{Akeson} R.,  et~al., 2019, \mn@doi [arXiv e-prints] {10.48550/arXiv.1902.05569}, \href {https://ui.adsabs.harvard.edu/abs/2019arXiv190205569A} {p. arXiv:1902.05569}

\bibitem[\protect\citeauthoryear{{Alcock} et~al.,}{{Alcock} et~al.}{1993}]{1993Natur.365..621A}
{Alcock} C.,  et~al., 1993, \mn@doi [\nat] {10.1038/365621a0}, \href {https://ui.adsabs.harvard.edu/abs/1993Natur.365..621A} {365, 621}

\bibitem[\protect\citeauthoryear{{Alcock} et~al.,}{{Alcock} et~al.}{1998}]{1998ApJ...499L...9A}
{Alcock} C.,  et~al., 1998, \mn@doi [\apjl] {10.1086/311355}, \href {https://ui.adsabs.harvard.edu/abs/1998ApJ...499L...9A} {499, L9}

\bibitem[\protect\citeauthoryear{{Aubourg} et~al.,}{{Aubourg} et~al.}{1993}]{1993Natur.365..623A}
{Aubourg} E.,  et~al., 1993, \mn@doi [\nat] {10.1038/365623a0}, \href {https://ui.adsabs.harvard.edu/abs/1993Natur.365..623A} {365, 623}

\bibitem[\protect\citeauthoryear{{Bhatiani}, {Dai}  \& {Guerras}}{{Bhatiani} et~al.}{2019}]{2019ApJ...885...77B}
{Bhatiani} S.,  {Dai} X.,   {Guerras} E.,  2019, \mn@doi [\apj] {10.3847/1538-4357/ab46ac}, \href {https://ui.adsabs.harvard.edu/abs/2019ApJ...885...77B} {885, 77}

\bibitem[\protect\citeauthoryear{{Binney} \& {Tremaine}}{{Binney} \& {Tremaine}}{2008}]{2008gady.book.....B}
{Binney} J.,  {Tremaine} S.,  2008, {Galactic Dynamics: Second Edition}

\bibitem[\protect\citeauthoryear{{Blaineau} et~al.,}{{Blaineau} et~al.}{2022}]{2022A&A...664A.106B}
{Blaineau} T.,  et~al., 2022, \mn@doi [\aap] {10.1051/0004-6361/202243430}, \href {https://ui.adsabs.harvard.edu/abs/2022A&A...664A.106B} {664, A106}

\bibitem[\protect\citeauthoryear{{Boveia} \& {Doglioni}}{{Boveia} \& {Doglioni}}{2018}]{2018ARNPS..68..429B}
{Boveia} A.,  {Doglioni} C.,  2018, \mn@doi [Annual Review of Nuclear and Particle Science] {10.1146/annurev-nucl-101917-021008}, \href {https://ui.adsabs.harvard.edu/abs/2018ARNPS..68..429B} {68, 429}

\bibitem[\protect\citeauthoryear{{Chartas}, {Krawczynski}, {Zalesky}, {Kochanek}, {Dai}, {Morgan}  \& {Mosquera}}{{Chartas} et~al.}{2017}]{2017ApJ...837...26C}
{Chartas} G.,  {Krawczynski} H.,  {Zalesky} L.,  {Kochanek} C.~S.,  {Dai} X.,  {Morgan} C.~W.,   {Mosquera} A.,  2017, \mn@doi [\apj] {10.3847/1538-4357/aa5d50}, \href {https://ui.adsabs.harvard.edu/abs/2017ApJ...837...26C} {837, 26}

\bibitem[\protect\citeauthoryear{{Corrigan} et~al.,}{{Corrigan} et~al.}{1991}]{1991AJ....102...34C}
{Corrigan} R.~T.,  et~al., 1991, \mn@doi [\aj] {10.1086/115856}, \href {https://ui.adsabs.harvard.edu/abs/1991AJ....102...34C} {102, 34}

\bibitem[\protect\citeauthoryear{{Dai} \& {Guerras}}{{Dai} \& {Guerras}}{2018}]{2018ApJ...853L..27D}
{Dai} X.,  {Guerras} E.,  2018, \mn@doi [\apjl] {10.3847/2041-8213/aaa5fb}, \href {https://ui.adsabs.harvard.edu/abs/2018ApJ...853L..27D} {853, L27}

\bibitem[\protect\citeauthoryear{{Dai}, {Chartas}, {Agol}, {Bautz}  \& {Garmire}}{{Dai} et~al.}{2003}]{2003ApJ...589..100D}
{Dai} X.,  {Chartas} G.,  {Agol} E.,  {Bautz} M.~W.,   {Garmire} G.~P.,  2003, \mn@doi [\apj] {10.1086/374548}, \href {https://ui.adsabs.harvard.edu/abs/2003ApJ...589..100D} {589, 100}

\bibitem[\protect\citeauthoryear{{Davis}, {Efstathiou}, {Frenk}  \& {White}}{{Davis} et~al.}{1985}]{1985ApJ...292..371D}
{Davis} M.,  {Efstathiou} G.,  {Frenk} C.~S.,   {White} S.~D.~M.,  1985, \mn@doi [\apj] {10.1086/163168}, \href {https://ui.adsabs.harvard.edu/abs/1985ApJ...292..371D} {292, 371}

\bibitem[\protect\citeauthoryear{{DeRocco}, {Smyth}  \& {Profumo}}{{DeRocco} et~al.}{2023}]{2023arXiv230813593D}
{DeRocco} W.,  {Smyth} N.,   {Profumo} S.,  2023, \mn@doi [arXiv e-prints] {10.48550/arXiv.2308.13593}, \href {https://ui.adsabs.harvard.edu/abs/2023arXiv230813593D} {p. arXiv:2308.13593}

\bibitem[\protect\citeauthoryear{{Draine}}{{Draine}}{1998}]{1998ApJ...509L..41D}
{Draine} B.~T.,  1998, \mn@doi [\apjl] {10.1086/311751}, \href {https://ui.adsabs.harvard.edu/abs/1998ApJ...509L..41D} {509, L41}

\bibitem[\protect\citeauthoryear{{Drake} \& {Cook}}{{Drake} \& {Cook}}{2003}]{2003ApJ...589..281D}
{Drake} A.~J.,  {Cook} K.~H.,  2003, \mn@doi [\apj] {10.1086/374640}, \href {https://ui.adsabs.harvard.edu/abs/2003ApJ...589..281D} {589, 281}

\bibitem[\protect\citeauthoryear{{Esteban-Guti{\'e}rrez}, {Mediavilla}, {Jim{\'e}nez-Vicente}  \& {Mu{\~n}oz}}{{Esteban-Guti{\'e}rrez} et~al.}{2023}]{2023ApJ...954..172E}
{Esteban-Guti{\'e}rrez} A.,  {Mediavilla} E.,  {Jim{\'e}nez-Vicente} J.,   {Mu{\~n}oz} J.~A.,  2023, \mn@doi [\apj] {10.3847/1538-4357/ace62f}, \href {https://ui.adsabs.harvard.edu/abs/2023ApJ...954..172E} {954, 172}

\bibitem[\protect\citeauthoryear{{Feng}}{{Feng}}{2010}]{2010ARA&A..48..495F}
{Feng} J.~L.,  2010, \mn@doi [\araa] {10.1146/annurev-astro-082708-101659}, \href {https://ui.adsabs.harvard.edu/abs/2010ARA&A..48..495F} {48, 495}

\bibitem[\protect\citeauthoryear{{Gerhard} \& {Silk}}{{Gerhard} \& {Silk}}{1996}]{1996ApJ...472...34G}
{Gerhard} O.,  {Silk} J.,  1996, \mn@doi [\apj] {10.1086/178039}, \href {https://ui.adsabs.harvard.edu/abs/1996ApJ...472...34G} {472, 34}

\bibitem[\protect\citeauthoryear{{Gott}}{{Gott}}{1981}]{1981ApJ...243..140G}
{Gott} J.~R. I.,  1981, \mn@doi [\apj] {10.1086/158576}, \href {https://ui.adsabs.harvard.edu/abs/1981ApJ...243..140G} {243, 140}

\bibitem[\protect\citeauthoryear{{Griest}, {Cieplak}  \& {Lehner}}{{Griest} et~al.}{2014}]{2014ApJ...786..158G}
{Griest} K.,  {Cieplak} A.~M.,   {Lehner} M.~J.,  2014, \mn@doi [\apj] {10.1088/0004-637X/786/2/158}, \href {https://ui.adsabs.harvard.edu/abs/2014ApJ...786..158G} {786, 158}

\bibitem[\protect\citeauthoryear{{Hawkins}}{{Hawkins}}{1993}]{1993Natur.366..242H}
{Hawkins} M.~R.~S.,  1993, \mn@doi [\nat] {10.1038/366242a0}, \href {https://ui.adsabs.harvard.edu/abs/1993Natur.366..242H} {366, 242}

\bibitem[\protect\citeauthoryear{{Hawkins}}{{Hawkins}}{1996}]{1996MNRAS.278..787H}
{Hawkins} M.~R.~S.,  1996, \mn@doi [\mnras] {10.1093/mnras/278.3.787}, \href {https://ui.adsabs.harvard.edu/abs/1996MNRAS.278..787H} {278, 787}

\bibitem[\protect\citeauthoryear{{Hawkins} \& {Veron}}{{Hawkins} \& {Veron}}{1993}]{1993MNRAS.260..202H}
{Hawkins} M.~R.~S.,  {Veron} P.,  1993, \mn@doi [\mnras] {10.1093/mnras/260.1.202}, \href {https://ui.adsabs.harvard.edu/abs/1993MNRAS.260..202H} {260, 202}

\bibitem[\protect\citeauthoryear{{Huchra}, {Gorenstein}, {Kent}, {Shapiro}, {Smith}, {Horine}  \& {Perley}}{{Huchra} et~al.}{1985}]{1985AJ.....90..691H}
{Huchra} J.,  {Gorenstein} M.,  {Kent} S.,  {Shapiro} I.,  {Smith} G.,  {Horine} E.,   {Perley} R.,  1985, \mn@doi [\aj] {10.1086/113777}, \href {https://ui.adsabs.harvard.edu/abs/1985AJ.....90..691H} {90, 691}

\bibitem[\protect\citeauthoryear{{Irwin}, {Webster}, {Hewett}, {Corrigan}  \& {Jedrzejewski}}{{Irwin} et~al.}{1989}]{1989AJ.....98.1989I}
{Irwin} M.~J.,  {Webster} R.~L.,  {Hewett} P.~C.,  {Corrigan} R.~T.,   {Jedrzejewski} R.~I.,  1989, \mn@doi [\aj] {10.1086/115272}, \href {https://ui.adsabs.harvard.edu/abs/1989AJ.....98.1989I} {98, 1989}

\bibitem[\protect\citeauthoryear{{Jim{\'e}nez-Vicente} \& {Mediavilla}}{{Jim{\'e}nez-Vicente} \& {Mediavilla}}{2022}]{2022ApJ...941...80J}
{Jim{\'e}nez-Vicente} J.,  {Mediavilla} E.,  2022, \mn@doi [\apj] {10.3847/1538-4357/ac9e59}, \href {https://ui.adsabs.harvard.edu/abs/2022ApJ...941...80J} {941, 80}

\bibitem[\protect\citeauthoryear{{Kerins}, {Binney}  \& {Silk}}{{Kerins} et~al.}{2002}]{2002MNRAS.332L..29K}
{Kerins} E.,  {Binney} J.,   {Silk} J.,  2002, \mn@doi [\mnras] {10.1046/j.1365-8711.2002.05463.x}, \href {https://ui.adsabs.harvard.edu/abs/2002MNRAS.332L..29K} {332, L29}

\bibitem[\protect\citeauthoryear{{Kochanek}}{{Kochanek}}{2004}]{2004ApJ...605...58K}
{Kochanek} C.~S.,  2004, \mn@doi [\apj] {10.1086/382180}, \href {https://ui.adsabs.harvard.edu/abs/2004ApJ...605...58K} {605, 58}

\bibitem[\protect\citeauthoryear{{Lemon} et~al.,}{{Lemon} et~al.}{2023}]{2023MNRAS.520.3305L}
{Lemon} C.,  et~al., 2023, \mn@doi [\mnras] {10.1093/mnras/stac3721}, \href {https://ui.adsabs.harvard.edu/abs/2023MNRAS.520.3305L} {520, 3305}

\bibitem[\protect\citeauthoryear{{Lewis} \& {Irwin}}{{Lewis} \& {Irwin}}{1995}]{1995MNRAS.276..103L}
{Lewis} G.~F.,  {Irwin} M.~J.,  1995, \mn@doi [\mnras] {10.1093/mnras/276.1.103}, \href {https://ui.adsabs.harvard.edu/abs/1995MNRAS.276..103L} {276, 103}

\bibitem[\protect\citeauthoryear{{McElwain} et~al.,}{{McElwain} et~al.}{2023}]{2023arXiv230101779M}
{McElwain} M.~W.,  et~al., 2023, arXiv e-prints, \href {https://ui.adsabs.harvard.edu/abs/2023arXiv230101779M} {p. arXiv:2301.01779}

\bibitem[\protect\citeauthoryear{{McGaugh}, {Schombert}, {Bothun}  \& {de Blok}}{{McGaugh} et~al.}{2000}]{2000ApJ...533L..99M}
{McGaugh} S.~S.,  {Schombert} J.~M.,  {Bothun} G.~D.,   {de Blok} W.~J.~G.,  2000, \mn@doi [\apjl] {10.1086/312628}, \href {https://ui.adsabs.harvard.edu/abs/2000ApJ...533L..99M} {533, L99}

\bibitem[\protect\citeauthoryear{{Millon} et~al.,}{{Millon} et~al.}{2020}]{2020A&A...640A.105M}
{Millon} M.,  et~al., 2020, \mn@doi [\aap] {10.1051/0004-6361/202037740}, \href {https://ui.adsabs.harvard.edu/abs/2020A&A...640A.105M} {640, A105}

\bibitem[\protect\citeauthoryear{{Mr{\'o}z} et~al.,}{{Mr{\'o}z} et~al.}{2017}]{2017Natur.548..183M}
{Mr{\'o}z} P.,  et~al., 2017, \mn@doi [\nat] {10.1038/nature23276}, \href {https://ui.adsabs.harvard.edu/abs/2017Natur.548..183M} {548, 183}

\bibitem[\protect\citeauthoryear{{Paczy\'nski}}{{Paczy\'nski}}{1986}]{1986ApJ...304....1P}
{Paczy\'nski} B.,  1986, \mn@doi [\apj] {10.1086/164140}, \href {https://ui.adsabs.harvard.edu/abs/1986ApJ...304....1P} {304, 1}

\bibitem[\protect\citeauthoryear{{Peacock}}{{Peacock}}{1999}]{1999coph.book.....P}
{Peacock} J.~A.,  1999, {Cosmological Physics}

\bibitem[\protect\citeauthoryear{{Poindexter} \& {Kochanek}}{{Poindexter} \& {Kochanek}}{2010}]{2010ApJ...712..658P}
{Poindexter} S.,  {Kochanek} C.~S.,  2010, \mn@doi [\apj] {10.1088/0004-637X/712/1/658}, \href {https://ui.adsabs.harvard.edu/abs/2010ApJ...712..658P} {712, 658}

\bibitem[\protect\citeauthoryear{{Rafikov} \& {Draine}}{{Rafikov} \& {Draine}}{2001}]{2001ApJ...547..207R}
{Rafikov} R.~R.,  {Draine} B.~T.,  2001, \mn@doi [\apj] {10.1086/318355}, \href {https://ui.adsabs.harvard.edu/abs/2001ApJ...547..207R} {547, 207}

\bibitem[\protect\citeauthoryear{{Schechter} et~al.,}{{Schechter} et~al.}{2003}]{2003ApJ...584..657S}
{Schechter} P.~L.,  et~al., 2003, \mn@doi [\apj] {10.1086/345716}, \href {https://ui.adsabs.harvard.edu/abs/2003ApJ...584..657S} {584, 657}

\bibitem[\protect\citeauthoryear{{Schechter}, {Wambsganss}  \& {Lewis}}{{Schechter} et~al.}{2004}]{2004ApJ...613...77S}
{Schechter} P.~L.,  {Wambsganss} J.,   {Lewis} G.~F.,  2004, \mn@doi [\apj] {10.1086/422907}, \href {https://ui.adsabs.harvard.edu/abs/2004ApJ...613...77S} {613, 77}

\bibitem[\protect\citeauthoryear{{Schild}}{{Schild}}{1996}]{1996ApJ...464..125S}
{Schild} R.~E.,  1996, \mn@doi [\apj] {10.1086/177304}, \href {https://ui.adsabs.harvard.edu/abs/1996ApJ...464..125S} {464, 125}

\bibitem[\protect\citeauthoryear{{Schmidt}, {Webster}  \& {Lewis}}{{Schmidt} et~al.}{1998}]{1998MNRAS.295..488S}
{Schmidt} R.,  {Webster} R.~L.,   {Lewis} G.~F.,  1998, \mn@doi [\mnras] {10.1046/j.1365-8711.1998.01326.x}, \href {https://ui.adsabs.harvard.edu/abs/1998MNRAS.295..488S} {295, 488}

\bibitem[\protect\citeauthoryear{{Schneider}}{{Schneider}}{1993}]{1993A&A...279....1S}
{Schneider} P.,  1993, \aap, \href {https://ui.adsabs.harvard.edu/abs/1993A&A...279....1S} {279, 1}

\bibitem[\protect\citeauthoryear{{Schneider} \& {Sluse}}{{Schneider} \& {Sluse}}{2013}]{2013A&A...559A..37S}
{Schneider} P.,  {Sluse} D.,  2013, \mn@doi [\aap] {10.1051/0004-6361/201321882}, \href {https://ui.adsabs.harvard.edu/abs/2013A&A...559A..37S} {559, A37}

\bibitem[\protect\citeauthoryear{{Sumi} et~al.,}{{Sumi} et~al.}{2011}]{2011Natur.473..349S}
{Sumi} T.,  et~al., 2011, \mn@doi [\nat] {10.1038/nature10092}, \href {https://ui.adsabs.harvard.edu/abs/2011Natur.473..349S} {473, 349}

\bibitem[\protect\citeauthoryear{{Sumi} et~al.,}{{Sumi} et~al.}{2023}]{2023AJ....166..108S}
{Sumi} T.,  et~al., 2023, \mn@doi [\aj] {10.3847/1538-3881/ace688}, \href {https://ui.adsabs.harvard.edu/abs/2023AJ....166..108S} {166, 108}

\bibitem[\protect\citeauthoryear{{Tisserand} et~al.,}{{Tisserand} et~al.}{2007}]{2007A&A...469..387T}
{Tisserand} P.,  et~al., 2007, \mn@doi [\aap] {10.1051/0004-6361:20066017}, \href {https://ui.adsabs.harvard.edu/abs/2007A&A...469..387T} {469, 387}

\bibitem[\protect\citeauthoryear{{Treu}}{{Treu}}{2010}]{2010ARA&A..48...87T}
{Treu} T.,  2010, \mn@doi [\araa] {10.1146/annurev-astro-081309-130924}, \href {https://ui.adsabs.harvard.edu/abs/2010ARA&A..48...87T} {48, 87}

\bibitem[\protect\citeauthoryear{{Trimble}}{{Trimble}}{1987}]{1987ARA&A..25..425T}
{Trimble} V.,  1987, \mn@doi [\araa] {10.1146/annurev.aa.25.090187.002233}, \href {https://ui.adsabs.harvard.edu/abs/1987ARA&A..25..425T} {25, 425}

\bibitem[\protect\citeauthoryear{{Tuntsov} \& {Walker}}{{Tuntsov} \& {Walker}}{2022}]{2022MNRAS.513.2491T}
{Tuntsov} A.~V.,  {Walker} M.~A.,  2022, \mn@doi [\mnras] {10.1093/mnras/stac998}, \href {https://ui.adsabs.harvard.edu/abs/2022MNRAS.513.2491T} {513, 2491}

\bibitem[\protect\citeauthoryear{{Tuntsov}, {Walker}  \& {Lewis}}{{Tuntsov} et~al.}{2004}]{2004MNRAS.352..125T}
{Tuntsov} A.~V.,  {Walker} M.~A.,   {Lewis} G.~F.,  2004, \mn@doi [\mnras] {10.1111/j.1365-2966.2004.07902.x}, \href {https://ui.adsabs.harvard.edu/abs/2004MNRAS.352..125T} {352, 125}

\bibitem[\protect\citeauthoryear{{Udalski}, {Szymanski}, {Kaluzny}, {Kubiak}, {Krzeminski}, {Mateo}, {Preston}  \& {Paczynski}}{{Udalski} et~al.}{1993}]{1993AcA....43..289U}
{Udalski} A.,  {Szymanski} M.,  {Kaluzny} J.,  {Kubiak} M.,  {Krzeminski} W.,  {Mateo} M.,  {Preston} G.~W.,   {Paczynski} B.,  1993, \actaa, \href {https://ui.adsabs.harvard.edu/abs/1993AcA....43..289U} {43, 289}

\bibitem[\protect\citeauthoryear{{Udalski} et~al.,}{{Udalski} et~al.}{2006}]{2006AcA....56..293U}
{Udalski} A.,  et~al., 2006, \actaa, \href {https://ui.adsabs.harvard.edu/abs/2006AcA....56..293U} {56, 293}

\bibitem[\protect\citeauthoryear{{Vakulik}, {Schild}, {Dudinov}, {Nuritdinov}, {Tsvetkova}, {Burkhonov}  \& {Akhunov}}{{Vakulik} et~al.}{2006}]{2006A&A...447..905V}
{Vakulik} V.,  {Schild} R.,  {Dudinov} V.,  {Nuritdinov} S.,  {Tsvetkova} V.,  {Burkhonov} O.,   {Akhunov} T.,  2006, \mn@doi [\aap] {10.1051/0004-6361:20053574}, \href {https://ui.adsabs.harvard.edu/abs/2006A&A...447..905V} {447, 905}

\bibitem[\protect\citeauthoryear{{Walker}}{{Walker}}{1999}]{1999MNRAS.308..551W}
{Walker} M.~A.,  1999, \mn@doi [\mnras] {10.1046/j.1365-8711.1999.02814.x}, \href {https://ui.adsabs.harvard.edu/abs/1999MNRAS.308..551W} {308, 551}

\bibitem[\protect\citeauthoryear{{Walker} \& {Lewis}}{{Walker} \& {Lewis}}{2003}]{2003ApJ...589..844W}
{Walker} M.~A.,  {Lewis} G.~F.,  2003, \mn@doi [\apj] {10.1086/374777}, \href {https://ui.adsabs.harvard.edu/abs/2003ApJ...589..844W} {589, 844}

\bibitem[\protect\citeauthoryear{{Walker} \& {Wardle}}{{Walker} \& {Wardle}}{2019}]{2019ApJ...881...69W}
{Walker} M.~A.,  {Wardle} M.~J.,  2019, \mn@doi [\apj] {10.3847/1538-4357/ab2987}, \href {https://ui.adsabs.harvard.edu/abs/2019ApJ...881...69W} {881, 69}

\bibitem[\protect\citeauthoryear{{Wambsganss}}{{Wambsganss}}{1992}]{1992ApJ...386...19W}
{Wambsganss} J.,  1992, \mn@doi [\apj] {10.1086/170987}, \href {https://ui.adsabs.harvard.edu/abs/1992ApJ...386...19W} {386, 19}

\bibitem[\protect\citeauthoryear{{Wambsganss}, {Schmidt}, {Colley}, {Kundi{\'c}}  \& {Turner}}{{Wambsganss} et~al.}{2000}]{2000A&A...362L..37W}
{Wambsganss} J.,  {Schmidt} R.~W.,  {Colley} W.,  {Kundi{\'c}} T.,   {Turner} E.~L.,  2000, \mn@doi [\aap] {10.48550/arXiv.astro-ph/0010232}, \href {https://ui.adsabs.harvard.edu/abs/2000A&A...362L..37W} {362, L37}

\bibitem[\protect\citeauthoryear{{Wertz} \& {Surdej}}{{Wertz} \& {Surdej}}{2014}]{2014MNRAS.442..428W}
{Wertz} O.,  {Surdej} J.,  2014, \mn@doi [\mnras] {10.1093/mnras/stu866}, \href {https://ui.adsabs.harvard.edu/abs/2014MNRAS.442..428W} {442, 428}

\bibitem[\protect\citeauthoryear{{Williams}}{{Williams}}{2020}]{2020RAA....20...44W}
{Williams} R.,  2020, \mn@doi [Research in Astronomy and Astrophysics] {10.1088/1674-4527/20/4/44}, \href {https://ui.adsabs.harvard.edu/abs/2020RAA....20...44W} {20, 044}

\bibitem[\protect\citeauthoryear{{Wo{\'z}niak}, {Alard}, {Udalski}, {Szyma{\'n}ski}, {Kubiak}, {Pietrzy{\'n}ski}  \& {Zebru{\'n}}}{{Wo{\'z}niak} et~al.}{2000}]{2000ApJ...529...88W}
{Wo{\'z}niak} P.~R.,  {Alard} C.,  {Udalski} A.,  {Szyma{\'n}ski} M.,  {Kubiak} M.,  {Pietrzy{\'n}ski} G.,   {Zebru{\'n}} K.,  2000, \mn@doi [\apj] {10.1086/308258}, \href {https://ui.adsabs.harvard.edu/abs/2000ApJ...529...88W} {529, 88}

\bibitem[\protect\citeauthoryear{{Wyithe} \& {Turner}}{{Wyithe} \& {Turner}}{2001}]{2001MNRAS.320...21W}
{Wyithe} J.~S.~B.,  {Turner} E.~L.,  2001, \mn@doi [\mnras] {10.1046/j.1365-8711.2001.03917.x}, \href {https://ui.adsabs.harvard.edu/abs/2001MNRAS.320...21W} {320, 21}

\bibitem[\protect\citeauthoryear{{Wyithe}, {Webster}  \& {Turner}}{{Wyithe} et~al.}{1999}]{1999MNRAS.309..261W}
{Wyithe} J.~S.~B.,  {Webster} R.~L.,   {Turner} E.~L.,  1999, \mn@doi [\mnras] {10.1046/j.1365-8711.1999.02844.x}, \href {https://ui.adsabs.harvard.edu/abs/1999MNRAS.309..261W} {309, 261}

\bibitem[\protect\citeauthoryear{{Wyithe}, {Webster}  \& {Turner}}{{Wyithe} et~al.}{2000a}]{2000MNRAS.312..843W}
{Wyithe} J.~S.~B.,  {Webster} R.~L.,   {Turner} E.~L.,  2000a, \mn@doi [\mnras] {10.1046/j.1365-8711.2000.03203.x}, \href {https://ui.adsabs.harvard.edu/abs/2000MNRAS.312..843W} {312, 843}

\bibitem[\protect\citeauthoryear{{Wyithe}, {Webster}  \& {Turner}}{{Wyithe} et~al.}{2000b}]{2000MNRAS.315...51W}
{Wyithe} J.~S.~B.,  {Webster} R.~L.,   {Turner} E.~L.,  2000b, \mn@doi [\mnras] {10.1046/j.1365-8711.2000.03360.x}, \href {https://ui.adsabs.harvard.edu/abs/2000MNRAS.315...51W} {315, 51}

\bibitem[\protect\citeauthoryear{{Wyrzykowski} et~al.,}{{Wyrzykowski} et~al.}{2011}]{2011MNRAS.413..493W}
{Wyrzykowski} {\L}.,  et~al., 2011, \mn@doi [\mnras] {10.1111/j.1365-2966.2010.18150.x}, \href {https://ui.adsabs.harvard.edu/abs/2011MNRAS.413..493W} {413, 493}

\bibitem[\protect\citeauthoryear{{Zackrisson} \& {Bergvall}}{{Zackrisson} \& {Bergvall}}{2003}]{2003A&A...399...23Z}
{Zackrisson} E.,  {Bergvall} N.,  2003, \mn@doi [\aap] {10.1051/0004-6361:20021762}, \href {https://ui.adsabs.harvard.edu/abs/2003A&A...399...23Z} {399, 23}

\bibitem[\protect\citeauthoryear{{Zackrisson}, {Bergvall}, {Marquart}  \& {Helbig}}{{Zackrisson} et~al.}{2003}]{2003A&A...408...17Z}
{Zackrisson} E.,  {Bergvall} N.,  {Marquart} T.,   {Helbig} P.,  2003, \mn@doi [\aap] {10.1051/0004-6361:20030895}, \href {https://ui.adsabs.harvard.edu/abs/2003A&A...408...17Z} {408, 17}

\bibitem[\protect\citeauthoryear{{Zwicky}}{{Zwicky}}{2009}]{2009GReGr..41..207Z}
{Zwicky} F.,  2009, \mn@doi [General Relativity and Gravitation] {10.1007/s10714-008-0707-4}, \href {https://ui.adsabs.harvard.edu/abs/2009GReGr..41..207Z} {41, 207}

\bibitem[\protect\citeauthoryear{{de Swart}, {Bertone}  \& {van Dongen}}{{de Swart} et~al.}{2017}]{2017NatAs...1E..59D}
{de Swart} J.~G.,  {Bertone} G.,   {van Dongen} J.,  2017, \mn@doi [Nature Astronomy] {10.1038/s41550-017-0059}, \href {https://ui.adsabs.harvard.edu/abs/2017NatAs...1E..59D} {1, 0059}

\bibitem[\protect\citeauthoryear{{van de Ven}, {Falc{\'o}n-Barroso}, {McDermid}, {Cappellari}, {Miller}  \& {de Zeeuw}}{{van de Ven} et~al.}{2010}]{2010ApJ...719.1481V}
{van de Ven} G.,  {Falc{\'o}n-Barroso} J.,  {McDermid} R.~M.,  {Cappellari} M.,  {Miller} B.~W.,   {de Zeeuw} P.~T.,  2010, \mn@doi [\apj] {10.1088/0004-637X/719/2/1481}, \href {https://ui.adsabs.harvard.edu/abs/2010ApJ...719.1481V} {719, 1481}

\makeatother
\end{thebibliography}

\appendix
\section{Magnification maps for images B and D}
\label{appendix:bdmaps}
\begin{figure*}
\centering
\includegraphics[width=160mm]{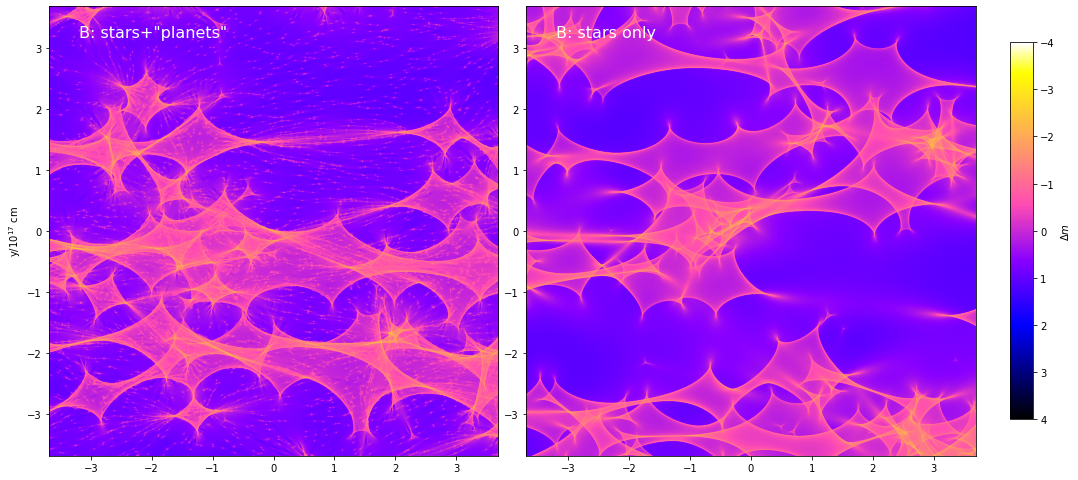}
\includegraphics[width=160mm]{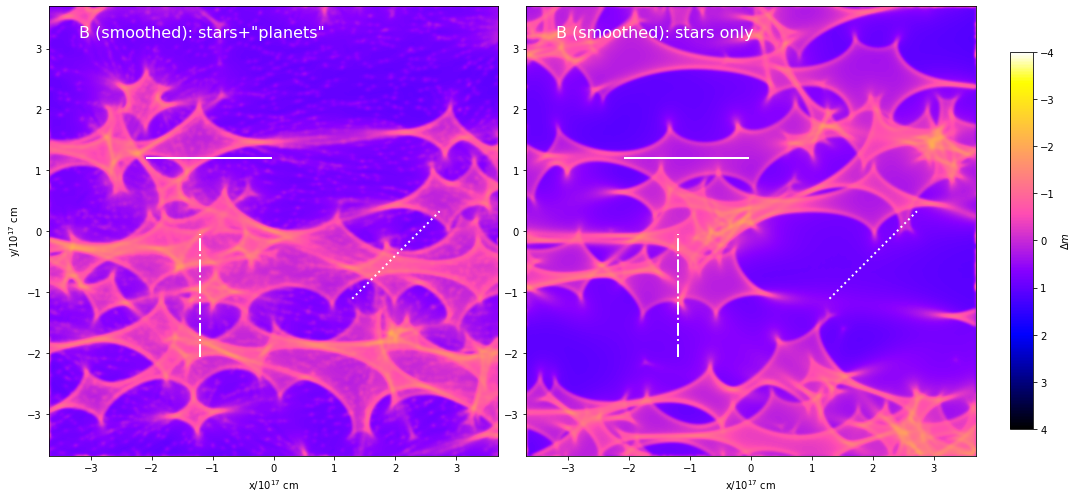}
\includegraphics[width=150mm]{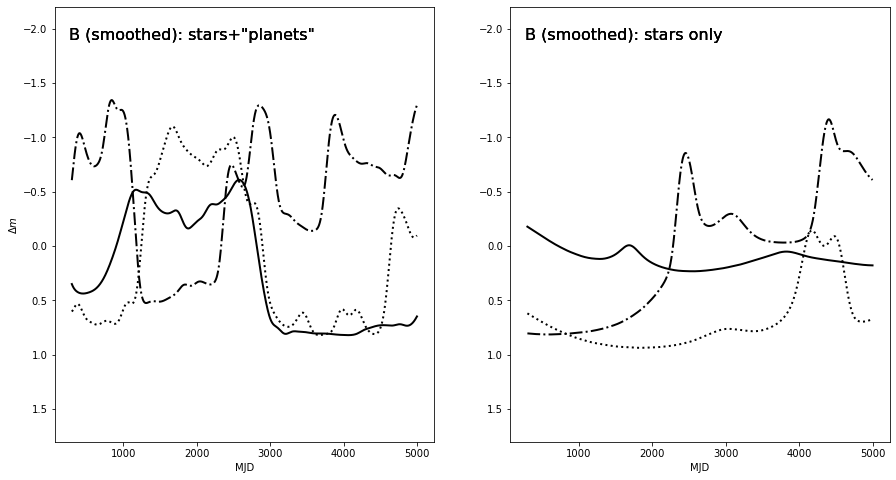}
\caption{Same as Figure~\ref{figure:magmapsA} for image B, of positive parity.}
    \label{figure:magmapsB}
\end{figure*}

\begin{figure*}
\centering
\includegraphics[width=160mm]{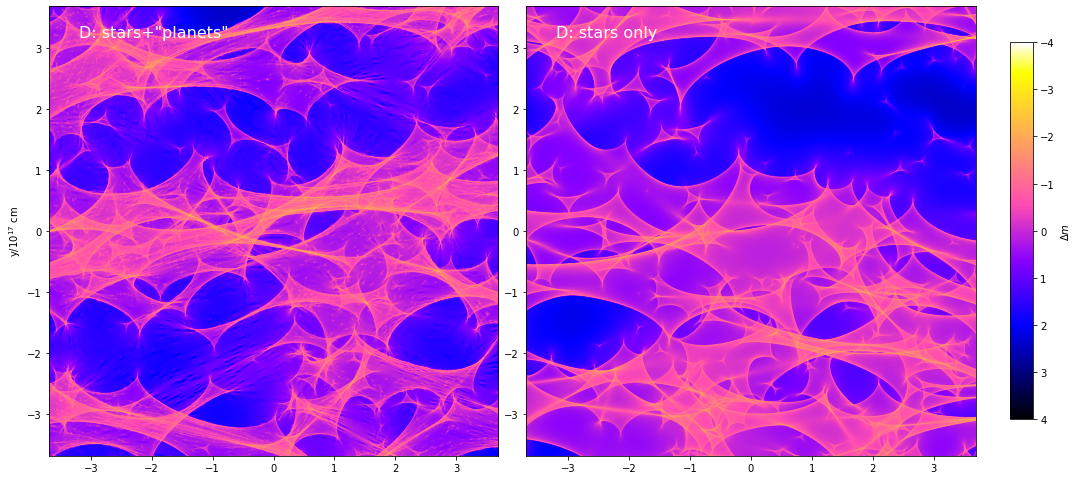}
\includegraphics[width=160mm]{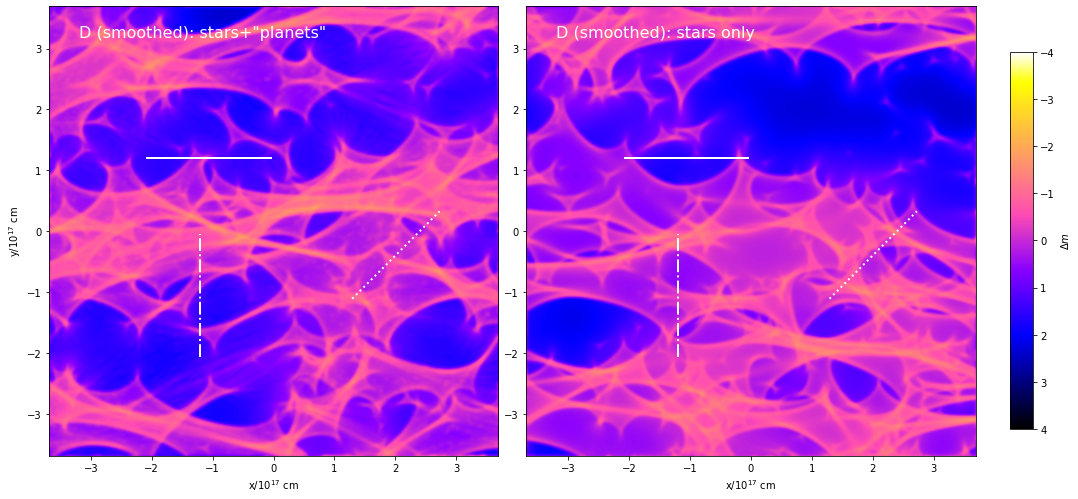}
\includegraphics[width=150mm]{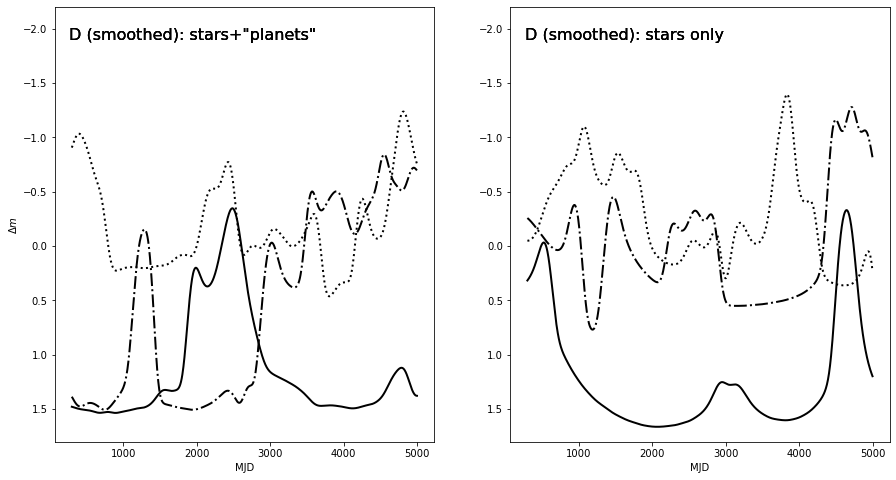}
\caption{Same as Figure~\ref{figure:magmapsA} for image D, of negative parity.}
    \label{figure:magmapsD}
\end{figure*}

Figures~\ref{figure:magmapsB} and~\ref{figure:magmapsD} present magnification maps and example light curves for images B and D; they are qualitatively similar to  Figures~\ref{figure:magmapsA} and~\ref{figure:magmapsC}.

\section{Relaxing source size and lens mass assumptions}
\label{appendix:sizenmass}
\begin{figure}
\centering
\includegraphics[width=70mm]{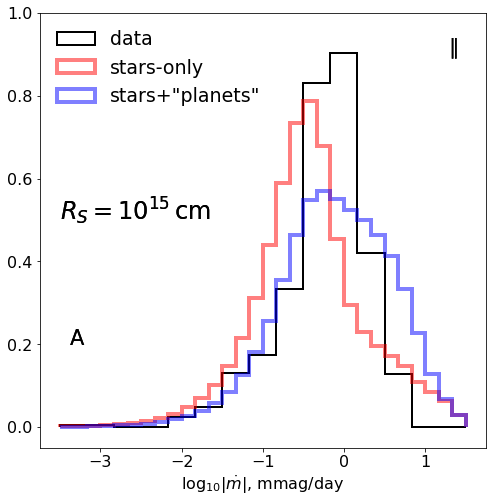}
\includegraphics[width=70mm]{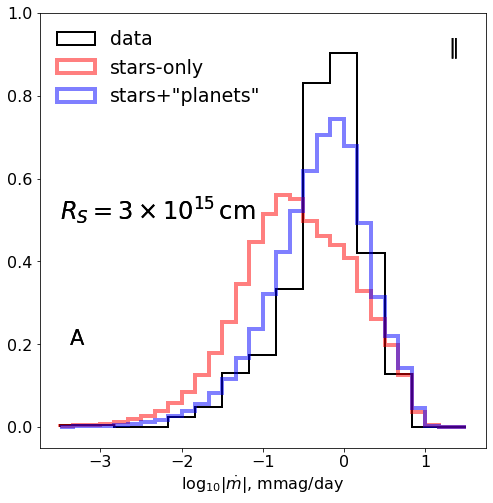}
\includegraphics[width=70mm]{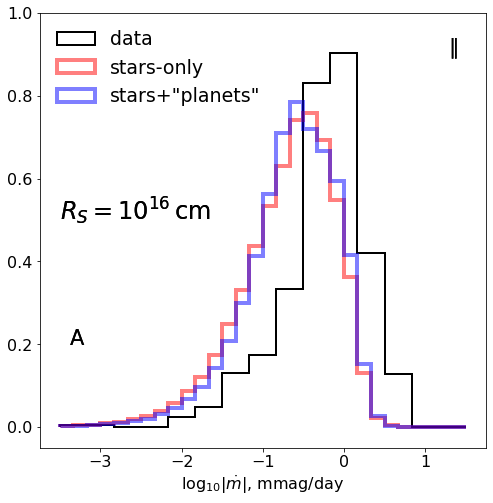}
\caption{An extended source suppresses the variations due to both stellar- and planetary-mass magnification, as summarised in Figure~\ref{figure:magstats}-like histograms for derivatives along the large-scale shear stretching directions of image A. In particular, a half light radius of just $R_S=10^{16}\,\mathrm{cm}$ totally desensitises this statistic to planetary-mass lenses in the \source\ case.}
    \label{figure:histsizeeffects}
\end{figure}

The nominal source half-light radius of $3\times10^{15}\,\mathrm{cm}$ chosen in this paper for \source\ is just over the Einstein radius of the planetary-mass lenses but is well below the Einstein scale of our stellar mass lenses, which reflects in the effect this smoothing has on the structures induced by two classes of lenses in Figures~\ref{figure:magmapsA}, \ref{figure:magmapsB}, \ref{figure:magmapsC} and~\ref{figure:magmapsD}. However, both types of structure induce the same (formally infinite) range of magnitude derivatives, and the effect of smoothing on the latter is not trivial. Figure~\ref{figure:histsizeeffects} shows how the statistics of light curve derivatives (for a representative image and track orientation) change as the source size is varied by half an order of magnitude either side of its fiducial value, the range likely exceeding the reasonable uncertainty of this estimate in the optical band. Smoothing clearly suppresses the derivative due to both stellar and planetary-mass lenses, but to a different extent, with a $10^{16}\,\mathrm{cm}$-wide source big enough to totally erase the nanolensing signal.

\begin{figure}
\centering
\includegraphics[width=\columnwidth]{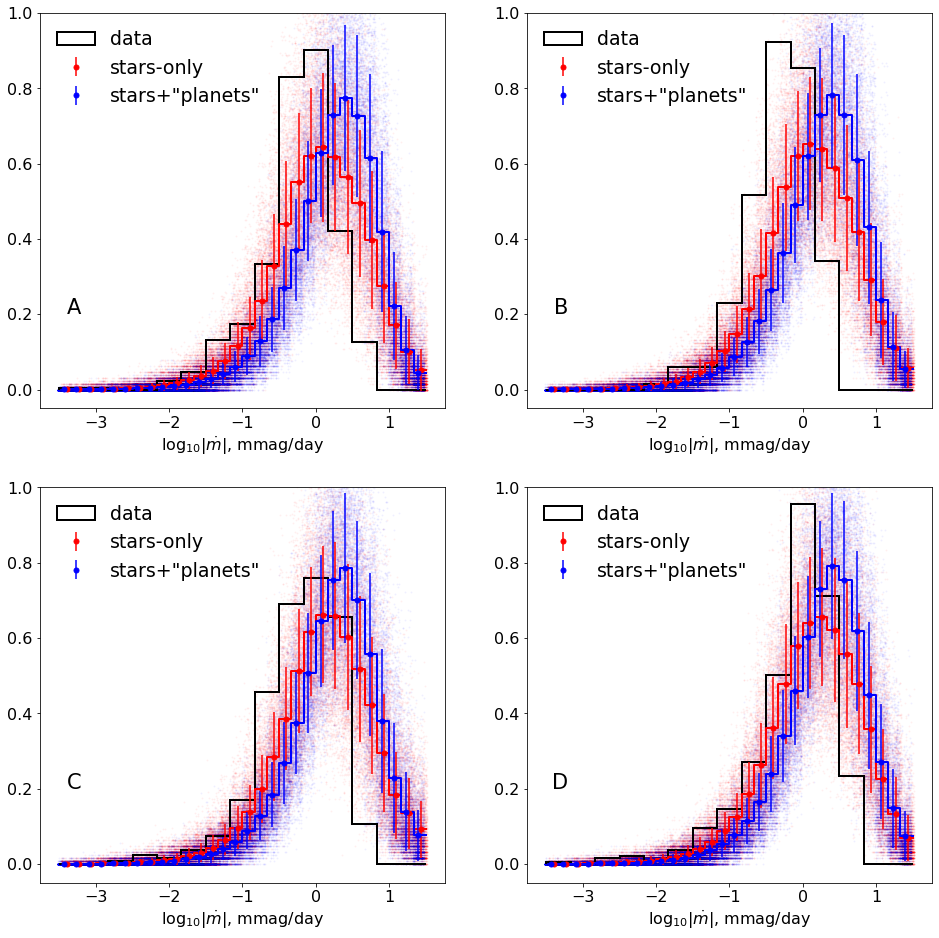}
\caption{Same as Figure~\ref{figure:datahistograms} assuming a source half-light radius of $R_S=10^{15}\,\mathrm{cm}$, half an order of magnitude smaller than used in Figure~\ref{figure:datahistograms}.}
    \label{figure:datahist15}
\end{figure}

\begin{figure}
\centering
\includegraphics[width=\columnwidth]{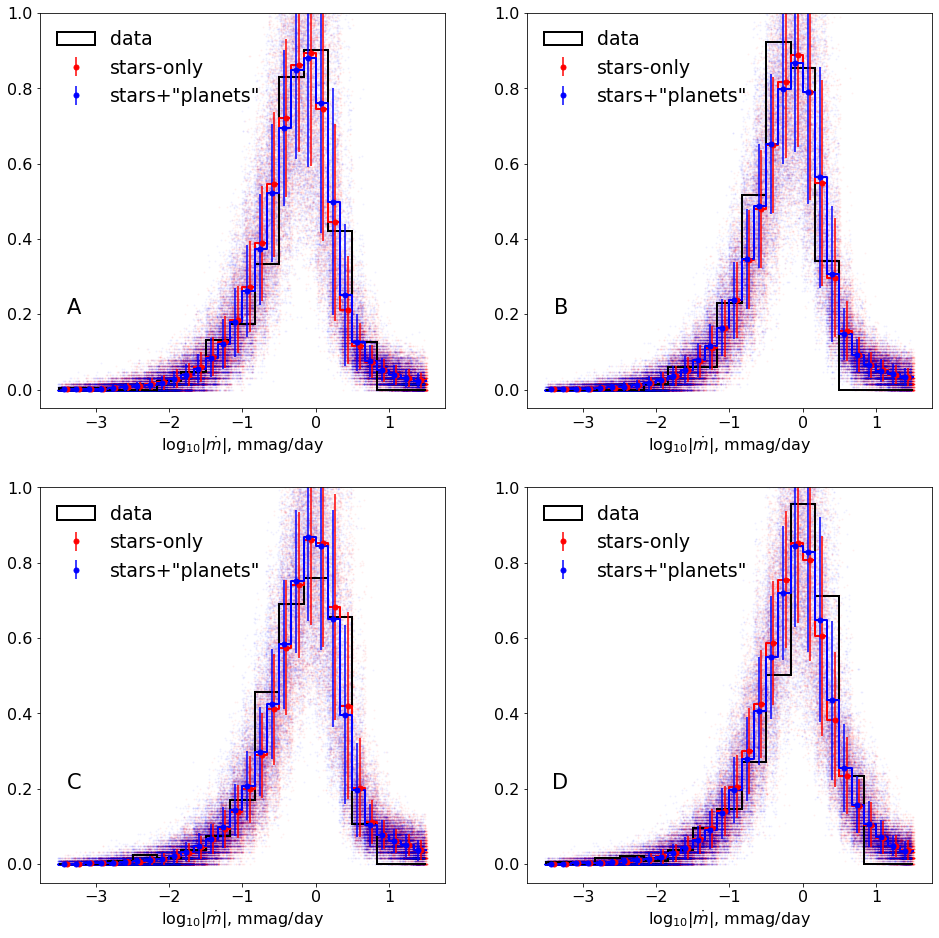}
\caption{Same as Figures~\ref{figure:datahistograms},\ref{figure:datahist15} for a larger source, $R_S=10^{16}\,\mathrm{cm}$.}
    \label{figure:datahist16}
\end{figure}

However, the source size effect is clearly subdominant to those of noise or irregular sampling, as revealed by Figures~\ref{figure:datahist15} and~~\ref{figure:datahist16}, showing analogues of Figure~\ref{figure:datahistograms} for smaller and larger source. Even in the case of the smallest source size that least suppresses the ``planetary'' signal, the limitations of the currently best available data (discussed in \S5) make it difficult to distinguish between the presence or absence of low-mass lenses. This reinforces the case for a better dataset, as presented in \S\ref{subsection:callfornewdata}.

From the discussion in this section one can also anticipate the effect of changing the nanolens mass to a smaller value. Reducing it to $10^{-5}\,\mathrm{M}_\odot$, would shrink the Einstein radius by half an order of magnitude, and so, at a constant mass density, would change the typical separation between the lenses. One would therefore expect that even a smaller source size, of $\sim10^{15}\,\mathrm{cm}$ would be quite detrimental to the discriminatory power of the light curve derivative statistics while the nanolensing signal would be strongly suppressed in the case of a half-light radius $R_S=3\times10^{15}\,\mathrm{cm}$ implied by the observed brightness of \source.

\begin{figure*}
\centering
\includegraphics[width=160mm]{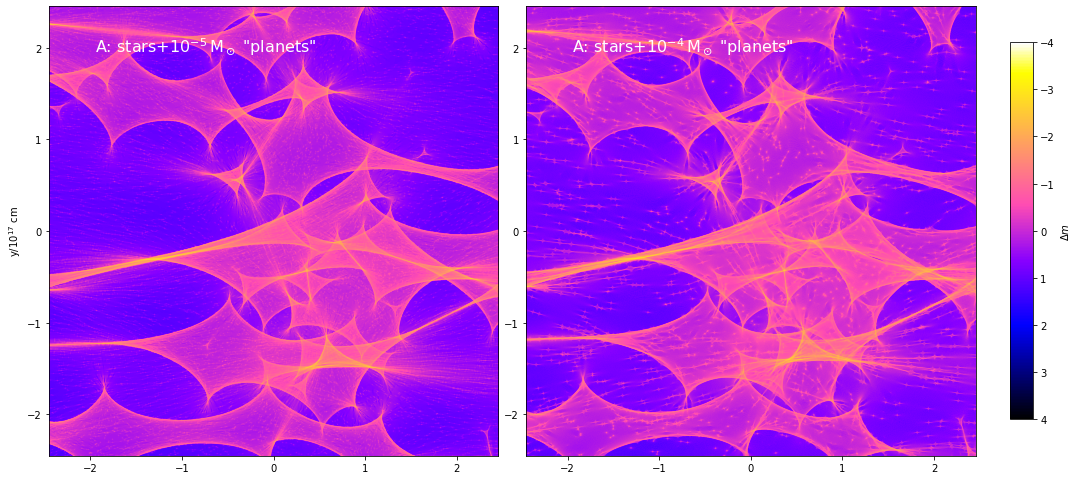}
\includegraphics[width=160mm]{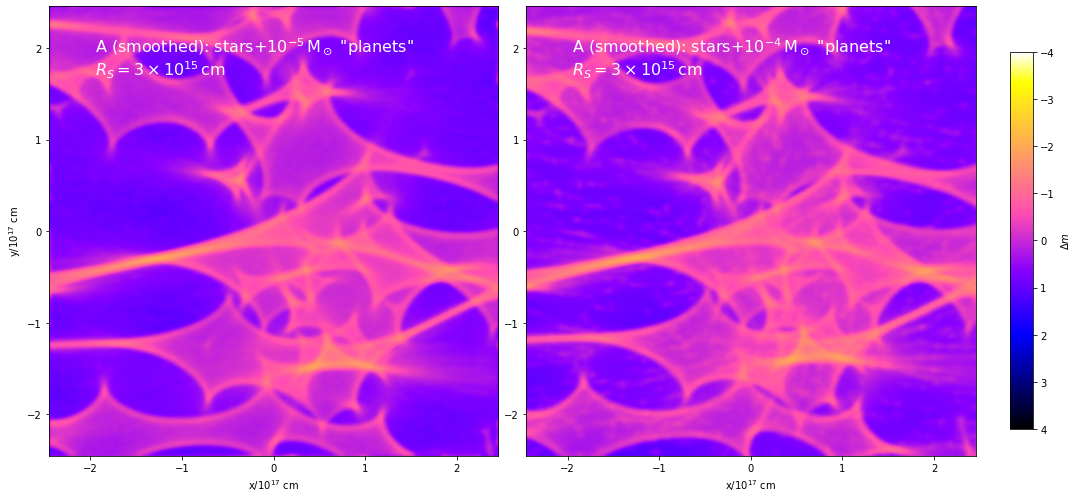}
\includegraphics[width=160mm]{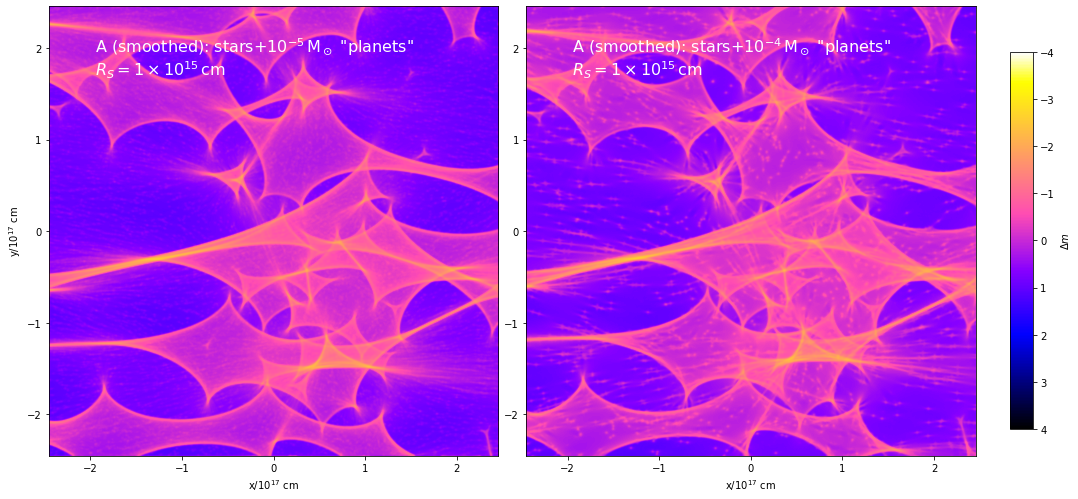}
\caption{Comparison of the effect of nano-lenses of a lower mass, $10^{-5}\,\mathrm{M}_\odot$ (left), to that of our fiducial $10^{-4}\,\mathrm{M}_\odot$ ``planets'' (right) for image A. The top row shows the original magnification maps, while the middle and bottom rows convolve them with a source of half-light radius $R_S=3\times 10^{15}\,\mathrm{cm}$ and $R_S=10^{15}\,\mathrm{cm}$.}
    \label{figure:magmapsminiA}
\end{figure*}

\begin{figure*}
\centering
\includegraphics[width=160mm]{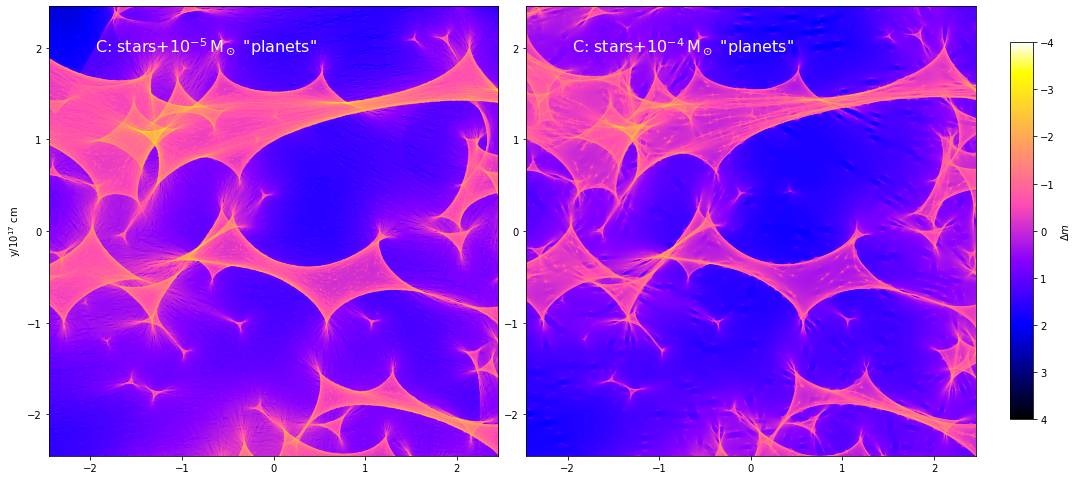}
\includegraphics[width=160mm]{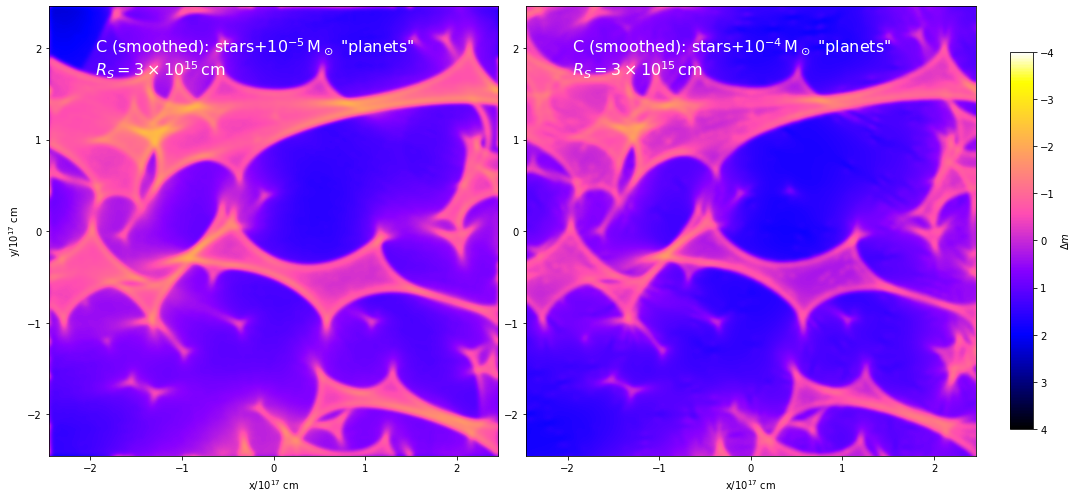}
\includegraphics[width=160mm]{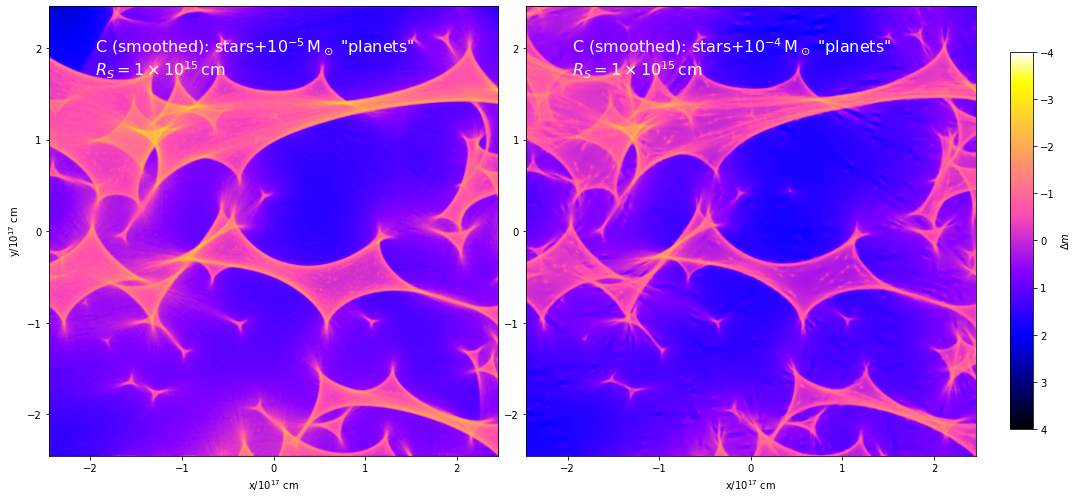}
\caption{Same as Figure~\ref{figure:magmapsminiA} for image C.}
    \label{figure:magmapsminiC}
\end{figure*}

Figures~\ref{figure:magmapsminiA},\ref{figure:magmapsminiC} illustrate those points. They present the magnification maps and their convolution with a Gaussian source of varying size assuming nano-lenses of different masses for two representative images of positive and negative parity (we kept the positions of the stellar mass lenses the same here). The magnification maps for these simulations were generated with the Fast Multipole Method based algorithm of \cite{2022ApJ...941...80J}, as the computational resources used for our simulations in Figures~\ref{figure:magmapsA}-\ref{figure:magmapsD} were no longer available. In order to properly resolve the Einstein radius of the lower mass lenses in this simulation, we had to decrease the pixel size -- which, given the limitation of the public interface\footnote{{\tt https://gloton.ugr.es/microlensing/}} to the \cite{2022ApJ...941...80J} software, necessitated simulating a slightly narrower field of view, of approximately $2.66\times10^{17}\,\mathrm{cm}$ on a side. We checked that the two codes agreed on the output when the resolution, field of view and the input positions and masses of the lenses are identical. The finer scale structure due to lower mass nano-lenses is severely degraded by smoothing with our fiducial source of half-light radius $R_S=3\times10^{15}\,\mathrm{cm}$ and the degradation is still strong even for $R_S=10^{15}\,\mathrm{cm}$.

\begin{figure}
\centering
\includegraphics[width=\columnwidth]{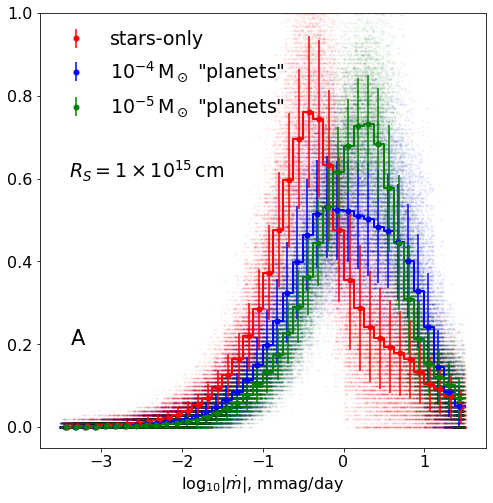}
\includegraphics[width=\columnwidth]{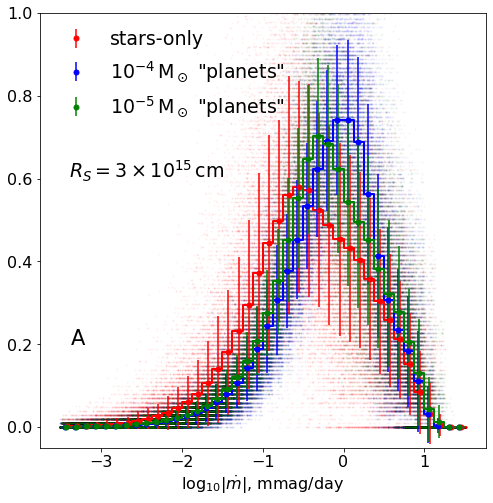}
\caption{Distribution of light curve derivatives for image A for different (if any) nanolens populations assuming a hypothetical dataset with weekly cadence and negligible noise. The top figure shows that in the case of a small source, $R_S=10^{15}\,\mathrm{cm}$, the statistics would even be sensitive to the mass of the ``planets'', although the bottom panel suggests that it would be difficult to go beyond establishing their very presence for a more realistic source size.}
    \label{figure:sizemasshist}
\end{figure}

Given this suppression, it is of little surprise that the light curve derivative statistics expected for lower mass lenses do not separate as well from the 'stars-only' case, further reducing the sensitivity to ``planets'' with the available data. Encouragingly though, with better sampling and increased accuracy, as advocated in Section~\ref{subsection:callfornewdata}, the statistics would be able to not only identify the presence of ``planets'' but also gauge their mass, as Figure~\ref{figure:sizemasshist} demonstrates in the case of a small source size. Only an indication of the presence of nano-lenses but not a determination of their mass would be possible for a more realistic value of $R_S$. The foregoing analysis suggests that even this quasar, which is particularly well-suited for the detection of low-mass microlenses by virtue of a very low lens-galaxy redshift, would be insensitive to the presence of still lower mass microlenses, for any reasonable estimate of the optical source size.

\section{Magnitude difference statistics}
\label{appendix:differences}
\begin{figure}
\centering
\includegraphics[width=\columnwidth]{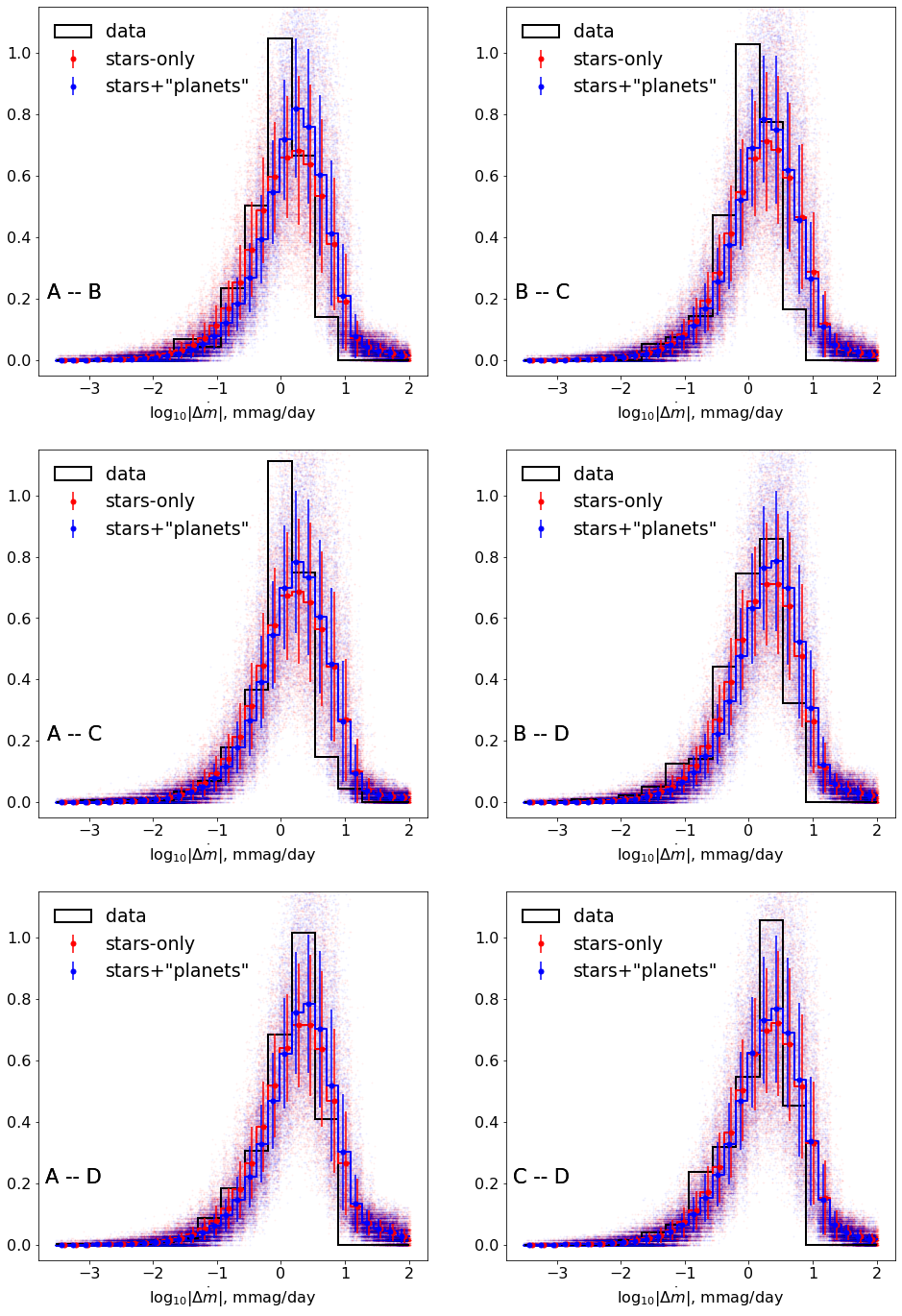}
\caption{Histograms of the (smoothed) difference light curve derivatives observed in all pairs of \source\ images compared to simulated light curves computed for magnification maps with and without planetary-mass microlenses, similar to that shown in Figure~\ref{figure:datahistograms}.}
    \label{figure:histdiff}
\label{lastpage}
\end{figure}

To avoid having to form an estimate the intrinsic variations of the source it is possible to construct pairwise magnitude differences, which automatically removes intrinsic variations when the time delay between images is negligible. Indeed that approach will be familiar to many readers and is a useful point of reference. We have therefore repeated our statistical analysis, but using pairwise magnitude differences, to see whether the data are better matched by simulated light-curves with or without the planetary-mass microlens population. The results, in the familiar form of magnitude (difference) derivative histograms, mimicking Figure~\ref{figure:datahistograms} are shown in Figure~\ref{figure:histdiff}. Unfortunately, none of the image pairs appears able to discriminate between the two hypotheses.

\end{document}